\documentclass{jfm}
\usepackage{graphicx}
\usepackage{epstopdf, epsfig}
\usepackage{amssymb}
\usepackage{amsmath}
\usepackage[dvipsnames]{xcolor}
\usepackage[]{lineno}

\shorttitle{Linearised RANS equations for the prediction of secondary flows}
\shortauthor{G. Zampino, D. Lasagna and B. Ganapathisubramani}

\title{Linearised Reynolds-Averaged predictions of secondary currents in turbulent channels with topographic heterogeneity}

\author{G. Zampino\aff{1}\corresp{\email{g.zampino@soton.ac.uk}},
  D. Lasagna\aff{1}
 \and B. Ganapathisubramani\aff{1}}

\affiliation{\aff{1}Aeronautics and Astronautics, Faculty of Engineering and Physical Sciences, University of Southampton, Hampshire, SO17 1BJ, UK}

\begin{document}
\maketitle

\begin{abstract}

A rapid predictive tool based on the linearised Reynolds-averaged Navier-Stokes equations is proposed in this work to investigate secondary currents generated by streamwise-independent surface topography modulations in turbulent channel flow. The tool is derived by coupling the Reynolds-averaged momentum equation to the Spalart-Allmaras transport equation for the turbulent eddy viscosity, using a nonlinear constitutive relation for the Reynolds stresses to capture correctly secondary motions. Linearised equations, describing the steady flow response to arbitrary surface modulations, are derived by assuming that surface modulations are shallow. Since the equations are linear, the superposition principle holds and the flow response induced by an arbitrary modulation can be obtained by combining appropriately the elementary responses obtained over sinusoidal modulations at multiple spanwise length scales. The tool permits a rapid exploration of large parameter spaces characterising structured surface topographies previously examined in the literature. Here, channels with sinusoidal walls and with longitudinal rectangular ridges are considered. For sinusoidal walls, a large response is observed at two spanwise wavelengths scaling in inner and outer units respectively, mirroring the amplification mechanisms in turbulent shear flows observed from transient growth analysis. For longitudinal rectangular ridges, the model suggests that the analysis of the response and the interpretation of the topology of secondary structures is facilitated when the ridge width and the gap between ridges are used instead of other combinations proposed in the literature.
\end{abstract}

\begin{keywords}
Linearised RANS equations, secondary currents, ridge-type roughness
\end{keywords}

\section{Introduction} 
When a wall-bounded turbulent flow develops over a surface with heterogeneous attributes, e.g. with lateral variations of the topography or of the roughness properties, secondary currents emerge in the form of coherent streamwise-aligned vortices. 
These flows, named by Prandtl secondary flows of the second kind \citep{prandtl1952}, have attracted significant interest since the first experiments in rectangular ducts with heterogeneous rough surfaces conducted by \citet{hinze1967,hinze1973}. In fact, these flows are highly relevant in many industrial and environmental applications, where aerodynamic surfaces are rarely smooth and homogeneous. Despite being relatively weak, with velocities of few percents the external velocity scale, these currents can alter natural wall-normal transport properties of wall-bounded turbulent flows \citep{volino2011, mejia2013,vanderwel2015,Hwanglee2018,medjnoun_vanderwel_ganapathisubramani_2020, zampiron2020} and can thus increase friction and heat transfer \citep{stroh2020},  and modify the performance of aerodynamic surfaces \citep{mejia2013, barros2014}. 

Broadly speaking, the heterogeneity can be distinguished between topographical variations, i.e. alternating regions of high/low relative elevation \citep{Hwanglee2018, medjnoun_vanderwel_ganapathisubramani_2018, medjnoun_vanderwel_ganapathisubramani_2020, castro2020}, and skin-friction variations, where the local wall shear stress varies as a consequence of changes in the surface attributes, such as the roughness properties \citep{barros2014, chung2018, stroh2019, forooghi} or over super-hydrophobic surfaces \citep{turk2014, stroh2016}. Combinations of these two have also been considered, (e.g \citet{vanderwel2015, yang2017, stroh2019}). However, in all cases, the flow topology observed above such surfaces is characterized by alternating high-momentum pathways (HMPs), corresponding to a downwash motion, and low-momentum pathways (LMPs), correlated to an upwash motion, as observed by \citet{mejia2013} and \citet{willigham2014}. This alternance of HMPs and LMPS is observed both experimentally \citep{barros2014,anderson2015,vanderwel2015} and numerically \citep{stroh2016,chung2018}. Even though the instantaneous field is highly complex \citep{vanderwel2019}, these motions are associated, in a Reynolds-averaged sense, to large scale streamwise vortical structures, driven by a turbulent torque produced by lateral variations of the (anisotropic) Reynolds stress tensor \citep{perkins1970, bottaro2006}.

The lateral organisation and intensity of HMPs and LMPs and of the associated vortical structures is often discussed in relation to a characteristic spanwise length scale of the heterogeneity, such as the spacing between longitudinal ridges or the width of roughness strips or patches of super-hydrophobic surface. Many authors have performed parametric studies and have demonstrated that secondary motions are most intense when this characteristic length scale is on the order of the thickness of the turbulent shear layer \citep{vanderwel2015, yang2017, chung2018}. However, significant changes in the flow topology,  {e.g the} appearance of tertiary flows, have also observed when other surface parameters are varied, such as the width of the ridges or the ridge geometry. 
In an effort to quantify these aspects, \citet{medjnoun_vanderwel_ganapathisubramani_2020} introduced the ratio between the cross-sectional areas above and below the mean surface height as the key surface parameter that distinguishes different topographies and the observed flow structure. They showed that the circulation of the time-averaged vortical structures is proportional to this ratio. However, a complete description of how surface characteristics influence the structure and intensity of secondary motions is still lacking. In fact, this endeavour has been hindered by the high-dimensional nature of the parameter space that characterises heterogeneous surfaces, which is costly to fully explore using experiments or scale-resolving simulations.

The overarching aim of this work is to develop a rapid predictive tool to aid the exploration of such spaces. In this paper, we restrict our attention to surfaces with lateral variations of the topography, but extensions to other types of heterogeneity are possible. The proposed tool is based on the steady linearised Reynolds-Averaged Navier-Stokes (RANS) equations, augmented by a turbulent eddy viscosity term. These equations have been used in past work to clarify key mechanisms of wall bounded turbulence. For instance, the characteristic spanwise length of near-wall streaks and large-scale motions in turbulent shear flows is well captured by the energy amplification properties of the Orr-Sommerfeld-Squire equations \citep{delalamo2006, pujals2009, hwangcossu2010}. \citet{luchini_charru_2010} and \citet{russoluchini2016} used linearised RANS equations to model flows over undulated bottoms or to examine the response to volume forcing. \citet{meyers2019} utilised the linearized RANS equations to predict the decay rate of dispersive stresses associated to secondary motions in the outer-layer region.  Unlike in some of the previous literature, where simple analytical profiles for the eddy viscosity have been used, here the Reynolds-averaged momentum equations is coupled with the Spalart-Allmaras transport equation for the turbulent eddy viscosity \citep{spalart1994}, to capture more faithfully the variable topography. Linearised equations are then derived by assuming that the topography is shallow when compared to any inner or outer length scale. For shallow modulations, the nonlinear convective terms are negligible and arbitrary surface topographies can be modelled using inhomogeneous linearised boundary conditions \citep{luchini2013}. Using these equations, the response of the  shear flow to an arbitrary, spectrally complex surface topography can be obtained by applying the superposition principle, i.e. by appropriately combining the elementary responses obtained for all the harmonic components defining the given surface. Channels with sinusoidal walls \citep{vidal2018} and with longitudinal rectangular ridges are considered in this paper as two paradigmatic configurations that have received significant attention in the recent literature.

The modelling technique and the linearisation of the governing equations is discussed in section \ref{methodology}. The approach is first applied to sinusoidal modulations in section \ref{sinusoidalwall}, to clarify the fundamental role of the spanwise length scale on the strength and structure of secondary motions. With this insight, channels with rectangular ridges are considered in section \ref{ridgeswall}. Finally, conclusions are reported in section \ref{conclusions}.

\section{Methodology} \label{methodology}

\subsection{Problem setup and equations of motion} \label{problemsetup}
\begin{figure}
    \centering
    \includegraphics[width=0.94\textwidth]{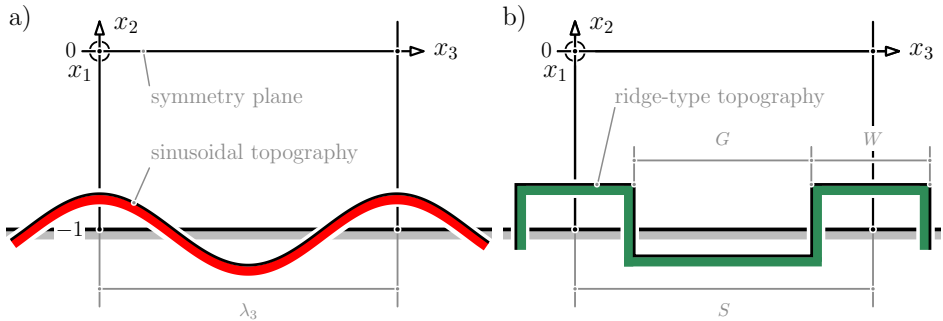}
    \caption{Sinusoidal, a), and ridge-type, b), topographies considered in this paper. The coordinate system $(x_1,x_2,x_3)$, with origin on the symmetry plane, is shown. The streamwise direction $x_1$ is oriented into the page. When scaled by $h$, the mean channel height is equal to $2$. Symmetric configurations obtained by mirroring the lower wall geometries shown in the diagrams about the mid-plane $x_2 = 0$ are considered. For sinusoidal topographies, the period of the modulation is denoted by $\lambda_3$. For ridge-type topographies, the spacing between elements (the period) is denoted by $S$, while $W$ and $G$ are used to indicate the ridge width and the gap between elements, respectively.}
    \label{sketch}
\end{figure}

The incompressible flow of a fluid with kinematic viscosity $\nu$ in a pressure-driven channel with fixed streamwise pressure gradient $\Pi$ is examined. The streamwise, wall-normal and spanwise directions, normalised by the channel mean half-height $h$, are identified by the Cartesian coordinates $(x_1, x_2, x_3)$, with the origin of the wall-normal coordinate located at the channel mid-plane. The friction velocity $u_\tau=\sqrt{\tau_w/\rho}$, with $\tau_w = h \Pi$ the mean wall friction, is used to normalize the velocity components $(u_1, u_2, u_3)$ along the three directions.
 {The reference pressure is $p_{ref}=\rho u_\tau^2$ and this leads to a nondimensional pressure gradient $\partial \bar{p}/\partial x_i = \delta_{i1}$, with $\delta_{ij}$ being the Kronocker delta}. Reynolds-averaging produces the mean velocity $\overline{u}_i$ and the fluctuation $u^\prime_i$. The superscript $(\cdot)^+$, generally used for inner scaled quantities, is omitted in the following to reduce clutter unless necessary.
With these definitions, the friction Reynolds number is $\Rey_\tau=u_\tau h/\nu$. We consider channels with streamwise-independent modulations of the wall topography, namely, sinusoidal modulations and rectangular ridges, as illustrated in figure \ref{sketch}.

The time-averaged flow structure in the channel is governed by the nondimensional Reynolds-averaged continuity and momentum equations
\begin{subequations}
\begin{align}
    \frac{\partial \overline{u}_i}{\partial x_i} =&\;0,\\
    \overline{u}_j\frac{\partial \overline{u}_i}{\partial x_j}=&- { \delta_{i1}}+\frac{1}{\Rey_\tau}\frac{\partial^2 \overline{u}_i}{\partial x_j^2}-\frac{\partial \overline{u_i'u_j'}}{\partial x_j},
   \label{RANS}
\end{align}
\end{subequations}
with no-slip boundary conditions on the two walls. As common, the trace of the Reynolds stress tensor is absorbed in the pressure term and we thus introduce the traceless stress tensor $\tau_{ij} = -\overline{u_i'u_j'} + \frac{1}{3}\overline{u_i'u_j'}\delta_{ij} $.
Assuming that a streamwise-independent mean flow (i.e. $\partial(\cdot)/\partial x_1 \equiv 0$) develops over streamwise-independent modulations, the mean pressure can be eliminated by employing
a streamwise velocity/streamfunction formulation, where the streamfunction $\overline{\psi}$ satisfies $\nabla^2 \overline{\psi}=\overline{\omega}_1$ with 
\begin{equation}
    \overline{\omega}_1= {\frac{\partial \overline{u}_3}{\partial x_2}-\frac{\partial \overline{u}_2}{\partial x_3}}
    \label{omegadefinition}
\end{equation} 
the streamwise vorticity.
With these definitions, the cross-stream velocity components are $\overline{u}_2=-{\partial \overline{\psi}}/{\partial x_3}$ and $\overline{u}_3={\partial \overline{\psi}}/{\partial x_2}$, satisfying automatically the continuity equation reduced to the cross-plane section.
The Reynolds-averaged streamwise momentum and streamfunction equations then become
\begin{subeqnarray}
    \displaystyle
  \frac{\partial \overline{\psi}}{ \partial x_2} \displaystyle \frac{\partial \overline{u}_1}{ \partial x_3} - && \displaystyle \frac{\partial \overline{\psi}}{ \partial x_3} \displaystyle \frac{\partial \overline{u}_1}{ \partial x_2} \!= \!1\!+\!\frac{1}{\Rey_\tau} \!\left( \displaystyle \frac{\partial^2 \overline{u}_1}{\partial x_2^2} \!+ \! \displaystyle \frac{\partial^2 \overline{u}_1}{\partial x_3^2}\!\right) \!+\! \displaystyle \frac{\partial \tau_{12}}{\partial x_2} \!+ \! \displaystyle \frac{\partial \tau_{13}}{\partial x_3}, \\
\displaystyle
  \nonumber  \displaystyle \frac{\partial^2}{\partial x_2 \partial x_3} \! && \!\left[ \! \left(\displaystyle \frac{\partial \overline{\psi}}{\partial x_2}\right)^2 -\left( \displaystyle \frac{\partial \overline{\psi}}{\partial x_3}\right)^2 \! \right] \! + \!\left( \displaystyle \frac{\partial^2}{\partial x_3^2} \!- \! \displaystyle \frac{\partial^2}{\partial x_2^2}\right)\!\displaystyle  \frac{\partial \overline{\psi}}{\partial x_2} \! \displaystyle \frac{\partial \overline{\psi}}{\partial x_3} = \,\;\,\;\, \\
  \displaystyle  \,\;\,\;\;\; \frac{1}{\Rey_\tau} \!&& \!\left( \displaystyle \frac{\partial^2}{\partial x_2^2} \!+\! \displaystyle \frac{\partial^2}{\partial x_3^2} \!\right)^2 \overline{\psi} 
  \!+\! \displaystyle \frac{\partial^2}{\partial x_2 \partial x_3} \left( \tau_{33}\!-\!\tau_{22}\right) \!+\! \left( \displaystyle \frac{\partial^2}{\partial x_2^2} \!-\! \displaystyle \frac{\partial^2}{\partial x_3^2} \!\right) \tau_{23}.
  \label{systemRANS}
\end{subeqnarray}

\subsection{Linearised response model} \label{linearmodel}
Without loss of generality, we assume the wall modulation to be spanwise-periodic, with fundamental period $\lambda_3$. We only consider zero-mean modulations of the wall geometry since perturbations of the mean channel height are trivially explained as a change in the Reynolds number, or as a wall-normal shift of the flow characteristics in boundary layers. Hence, an arbitrary modulation can be expressed by a function $f(x_3)$, with cosine series
\begin{equation}\label{fx3}
    f(x_3) = \sum_{n=1}^{\infty} f^n \cos(n k_3 x_3),
\end{equation}
with $k_3=2 \pi/\lambda_3$ the fundamental wavenumber  and $f^n$ the amplitude of the n-th wavenumber mode. Expressions for $f(x_3)$ for the two surfaces considered in the present work are given in equations \eqref{eqn:sinusoidal} and \eqref{eq:ridge_geometry}, respectively. Following \citet{russoluchini2016}, we then assume that the amplitude of the modulation is smaller than any other relevant geometric or flow length scale and we introduce a small parameter $\epsilon \ll 1$. The lower channel wall is then located at $x_2 = - 1 + \epsilon  f(x_3)$, while several configurations are possible for the upper wall.  {In one of the alternatives, the upper wall is located at $x_2 = 1 - \epsilon f(x_3)$, defining symmetric channels where secondary currents occupy at most half the channel height. In anti-symmetric channels, where the upper wall is located at $x_2 = 1 + \epsilon f(x_3)$, as in \citet{vidal2018}, the secondary currents can occupy the entire channel height and interact with the two shear layers developing over the top and bottom walls. In order to model more closely secondary currents in boundary layers \citep{Hwanglee2018, vanderwel2015, medjnoun_vanderwel_ganapathisubramani_2020} or open channel flows \citep{zampiron2020}, where the secondary currents develop across the shear flow, symmetric channels are considered in this paper.}

In a small-modulation scenario, a generic time-averaged quantity ${q}(x_2, x_3)$ in the channel with modulated walls (dropping the overbar to reduce clutter) can be expanded in a Taylor series in $\epsilon$ as
\begin{equation}
    q(x_2, x_3)=q^{(0)}(x_2)+\epsilon {q}^{(1)}(x_2, x_3)+ \mathcal{O}(\epsilon^2),
    \label{lineardecomposition}
\end{equation}
where ${q}^{(0)}$ denotes the plane channel solution. This expansion implies that the strength of secondary flows produced by a shallow modulation varies linearly with the amplitude $\epsilon$ and the perturbation quantity ${q}^{(1)}$ can be thus interpreted as the flow response (i.e. secondary currents) for a unitary change of the wall geometry given by \eqref{fx3}.

Substituting the Taylor expansion (\ref{lineardecomposition}) for all flow variables in the Reynolds-averaged equations \eqref{systemRANS} and considering terms at order zero in $\epsilon$, the time-averaged streamwise momentum equation is
\begin{equation}
\displaystyle
0=1 + \displaystyle\frac{1}{\Rey_\tau} \frac{\partial^2 u_1^{(0)}}{\partial x_2^2} + \frac{\partial \tau_{12}^{(0)}}{\partial x_2}, 
\label{systemzeroorder}
\end{equation}
while the streamfunction equation is trivially satisfied, since $u^{(0)}_2 = u^{(0)}_3 = 0$ in a plane channel. Retaining terms at order one in $\epsilon$, we obtain the set of equations
\begin{subeqnarray}\label{eq:linear_equations}
\displaystyle
\hspace{-0.8cm}-\frac{\partial \psi^{(1)}}{\partial x_3} \Gamma &&\!=\!\frac{1}{\Rey_\tau}\!\left( \frac{\partial^2 }{\partial x_2^2}\!+\!\frac{\partial^2 }{\partial x_3^2}\right) u_1^{(1)}\!+\!\frac{\partial \tau_{12}^{(1)} }{\partial x_2}\!+\!\frac{\partial \tau_{13}^{(1)} }{\partial x_3},
    \label{OSfinal}\\
    \displaystyle
  0&&\!=\!\frac{1}{\Rey_{\tau}}\!\left( \frac{\partial^2 }{\partial x_2^2}\!+\!\frac{\partial^2}{\partial x_3^2}\right)^2\! \psi^{(1)}\!+\!\frac{\partial^2}{\partial x_2 \partial x_3} \left( \tau_{33}^{(1)}\!-\!\tau_{22}^{(1)}\right)\!+\!\left( \frac{\partial^2}{\partial x_2^2}\!-\!\frac{\partial^2}{\partial x_3^2}\right)\! \tau_{23}^{(1)},
\label{psiequation}
\end{subeqnarray}
where $\Gamma=\partial u_1^{(0)}/\partial x_2$. These equations describe the new equilibrium between the perturbation of mean flow quantities ($u_1^{(1)}$, $\psi^{(1)}$) and the perturbation of the turbulent stress tensor $\tau_{ij}^{(1)}$. It is worth pointing out that the term ${\partial \psi^{(1)}}/ {\partial x_3} \Gamma$, analogous to the off-diagonal coupling operator in the Orr-Sommerfeld-Squire linearised equations \citep{schmid2000stability}, is the only coupling term explicitly appearing in this set of equations. Physically, this terms produces a spanwise modulation of the streamwise velocity as a result of secondary motions in the cross-stream plane.

The key property of these equations is linearity, since second order perturbation-perturbation terms arising from the convective nonlinearity are neglected at order one.  
As pointed out in \cite{meyers2019}, neglecting these terms is justified by the fact that the cross-stream velocity components are generally quite weak, i.e.~less than 5\% the external velocity scale \citep{anderson2015, medjnoun_vanderwel_ganapathisubramani_2020, Hwanglee2018}, especially at large distances from the wall. The key advantage is that the flow response induced by an arbitrary, spectrally-complex modulation $f(x_3)$ can be obtained by appropriately combining solutions of linear equations obtained at each spanwise wavenumber characterising the modulation in the expansion (\ref{fx3}).

\subsection{Nonlinear Reynolds stress model}
\label{sec:turbulence_closure_model}
To close the mean equations at order zero and one, it is now necessary to express the Reynolds stress tensor as a function of other mean quantities. One option is to introduce a linear Boussinesq hypothesis, using the turbulent eddy viscosity $\nu_t$ to derive the linear constitutive relation
\begin{equation}
\tau_{ij}^L= 2 \nu_t S_{ij} \label{eqn:taulindefinition}
\end{equation}
with $S_{ij}$ the mean velocity gradient tensor 
 \begin{equation}
    S_{ij}=\displaystyle \frac{1}{2} \left( \displaystyle \frac{\partial \overline{u}_i}{ \partial x_j}+ \displaystyle \frac{\partial \overline{u}_j}{\partial x_i}\right).
\end{equation}  
Expanding the turbulent stresses in a Taylor series as in (\ref{lineardecomposition}), the leading terms at order zero and one are
\begin{eqnarray}
    \tau_{ij}^{L(0)}&=&2 \nu_t^{(0)} S_{ij}^{(0)}, \\
    \tau_{ij}^{L(1)}&=&2 \nu_t^{(0)} S_{ij}^{(1)}+2 \nu_t^{(1)} S_{ij}^{(0)}.
\end{eqnarray}
where $\nu_t^{(1)}$ is the unknown perturbation of the eddy viscosity profile induced by the wall modulation. When a linear relation is used, however, no secondary flows are predicted \citep{perkins1970, speziale1982, bottaro2006}. In fact, the stresses appearing in  {(\ref{psiequation}b)} would not depend on the streamwise velocity since the stress tensor is isotropic and the streamfunction equation  {(\ref{psiequation}b)} decouples from the streamwise momentum equation  {(\ref{psiequation}a)}. Transient energy amplification from inhomogeneous initial conditions can be observed \citep{delalamo2006, pujals2009} but the steady response to an exogenous forcing, e.g. from the wall modulation, is trivial, $\psi^{(1)} \equiv 0$. 
Hence, a nonlinear Reynolds stress model is necessary. Several approaches have been described in the literature (e.g. \citet{speziale1991,speziale1991b, liencubic}).
Here we use the Quadratic Constitutive Relation (QCR) nonlinear model introduced by \citet{spalart2000}, which contains simple terms proportional to the product of the rotation and the strain tensors. This model was recently utilised by \citet{spalart2018} to predict the high-Reynolds number asymptotic properties of secondary flows in square and elliptical ducts, providing a good approximation of the secondary vortical flow topology and of the wall friction coefficient. Compared to other approaches, the QCR model is straightforward to manipulate analytically, and it is thus chosen here to remain in the original spirit of developing a simple predictive model of secondary flows over heterogeneous surfaces. 

In the QCR model, the Reynolds stresses become
\begin{equation}
    \tau_{ij}=\tau_{ij}^{L}-C_{r1}\left[ O_{ik}\tau_{jk}^{L}+O_{jk}\tau_{ik}^{L}\right], \label{taunonlinear}
\end{equation}
where the tuning constant $C_{r1}$ controls the anisotropy of the Reynolds stress tensor. \citet{spalart2000} suggests using $C_{r1}=0.3$ to match the anisotropy in the outer region of wall-bounded turbulent flows and we follow this indication in this paper.
In \eqref{taunonlinear}, $O_{ij}$ is the normalised rotation tensor
\begin{equation}
O_{ij}  = \frac{2W_{ij}}{\sqrt{\displaystyle  {\frac{\partial \overline{u}_m}{\partial x_n} \frac{\partial \overline{u}_m}{\partial x_n}}}}, \quad \mathrm{with}\quad
W_{ij} = \frac{1}{2}\left( \frac{\partial \overline{u}_i}{\partial x_j}-\frac{\partial \overline{u}_j}{\partial x_i}\right).
\label{oijdefinition}
\end{equation}
 
At order zero, the nonlinear stress tensor is equal to the expression obtained from the linear constitutive relation. At first-order, the Reynolds stress tensor is 
\begin{equation}\label{firstordertau}
   	\tau_{ij}^{(1)} =  \tau_{ij}^{L(1)}
   	 - C_{r1}\!\left[O_{ik}^{(1)}\tau_{jk}^{L(0)} + O_{ik}^{(0)}\tau_{jk}^{L(1)} + O_{jk}^{(1)}\tau_{ik}^{L(0)} + O_{jk}^{(0)}\tau_{ik}^{L(1)}\right],
\end{equation}
where $O_{ij}^{(1)}$ is the normalised rotation tensor induced by the first-order velocity components (see appendix \ref{appendix:qcr}). 
Developing (\ref{firstordertau}), the individual perturbation Reynolds stresses appearing in \eqref{eq:linear_equations} are
\begin{subeqnarray}
\label{tau23qcr}
    	\tau_{12}^{(1)}&=&\nu_t^{(0)}\displaystyle \frac{\partial u_1^{(1)}}{\partial x_2}+ \nu_t^{(1)}\Gamma+2C_{r1} \mathrm{sign}(\Gamma)\nu_t^{(0)} \displaystyle \frac{\partial^2 \psi^{(1)}}{\partial x_2 \partial x_3},\\
	\tau_{13}^{(1)}&=&\nu_t^{(0)} \displaystyle \frac{\partial u_1^{(1)}}{\partial x_3}-2C_{r1} \mathrm{sign}(\Gamma)\nu_t^{(0)} \frac{\partial^2 \psi^{(1)}}{\partial x_2^2},\\
	\tau_{23}^{(1)}&=&\nu_t^{(0)}\left(\displaystyle \frac{\partial^2}{\partial x_2^2}- \frac{\partial^2}{\partial x_3^2}\right) \psi^{(1)}+2 C_{r1} \mathrm{sign}(\Gamma) \nu_t^{(0)} \frac{\partial u_{1}^{(1)}}{\partial x_3},\\
	\tau_{22}^{(1)}&=&-2 \nu_t^{(0)} \displaystyle \frac{\partial^2 \psi^{(1)}}{\partial x_2 \partial x_3}+2 C_{r1} \left[ \mathrm{sign}(\Gamma)\nu_t^{(0)} \displaystyle \frac{\partial u_{1}^{(1)}}{\partial x_2}+\mathrm{sign}(\Gamma) \nu_t^{(1)}\Gamma\right],\\
	\tau_{33}^{(1)}&=&2 \nu_t^{(0)} \displaystyle \frac{\partial^2 \psi^{(1)}}{\partial x_2 \partial x_3},
\end{subeqnarray}
where `$\mathrm{sign}$' is the sign function. Except for $\tau_{33}^{(1)}$, which coincides with its linear Boussinesq's definition, all other stresses contain an additional term specific to the QCR model, which results in a tighter, two-way coupling between the streamfunction and streamwise velocity equations, able to sustain secondary currents.

\subsection{Eddy viscosity transport model}
The perturbation of the turbulent stresses \eqref{tau23qcr} still contains the unknown perturbation eddy viscosity $\nu_t^{(1)}$.
Past studies that have utilised linearised RANS equations to examine transient energy amplification in plane turbulent channels \citep{delalamo2006, pujals2009} have often used analytical eddy-viscosity profiles \citep{cess1958, reynolds1972}.
In these works, the eddy viscosity was assumed to be constant and not influenced by the growth of the optimal structures. This assumption, however, has little physical justification for a modulated geometry.
To provide a better description of the eddy viscosity distribution in the modulated geometry and capture transport effects, we use 
in the present paper the one-equation Spalart-Allmaras (SA) turbulence transport model \citep{spalart1994}, initially developed for attached shear flows. 
Using the channel half-height and the friction velocity for normalisation, the SA model introduces one transport equation for the transformed eddy viscosity $\tilde{\nu}$ related to the turbulent viscosity by the relation
\begin{equation}
    \nu_t=\tilde{\nu} f_{v1},
\end{equation}
where
\begin{equation}\label{eq:fv1_definition}
    f_{v1}=\frac{\chi^3}{\chi^3+c_{v1}^3},
\end{equation} 
with $\chi= \Rey_\tau \tilde{\nu}$ and $c_{v1}$ a tuning constant. The modified eddy viscosity coincides with the turbulent viscosity away from the wall. Additionally, the term \eqref{eq:fv1_definition} ensures the correct decay of the turbulent viscosity in the viscous sublayer \citep{spalart1994, mellor1968} when $\tilde{\nu}$ behaves linearly in the log layer down to the surface, which is advantageous for numerical reasons.
The steady transport equation for $\tilde{\nu}$,
\begin{equation}
        \overline{u}_i \frac{\partial \tilde{\nu}}{\partial x_i} =c_{b1} \tilde{\mathcal{S}} \tilde{\nu}+\frac{1}{\sigma}\left\{ \frac{\partial }{\partial x_j}\left[ \left( \frac{1}{\Rey_\tau}+\tilde{\nu}\right) \frac{\partial \tilde{\nu} }{\partial x_j }\right] + c_{b2} \frac{\partial \tilde{\nu}}{ \partial x_j}\frac{\partial \tilde{\nu}}{ \partial x_j}  \right\}-c_{w1} f_w \left( \frac{\tilde{\nu}}{d}\right)^2,
    \label{SAequation1}
\end{equation}
is composed by a convection, production, diffusion and destruction terms. In the production term, the quantity $\tilde{\mathcal{S}}$ is defined as
\begin{equation}
 \tilde{\mathcal{S}}=\sqrt{2W_{ij} W_{ij}}+ \displaystyle \frac{\tilde{\nu}}{k^2 d^2} f_{v2} \quad \mathrm{with}  \quad f_{v2}= 1- \displaystyle \frac{\chi}{1+ \chi f_{v1}}.
\end{equation}
with $k$ the von K\'arm\'an constant. The destruction term in (\ref{SAequation1}) captures the blocking effect of the wall on turbulent fluctuations and is a function of the distance to the nearest surface $d$. With this term, the model produces an accurate log-layer in wall-bounded flows. It includes
a nondimensional function $f_{w}$  that increases the decay of the destruction term in the outer region. This term reads as
\begin{equation}
    f_{w} = g \left[ \displaystyle \frac{1+ c_{w3}^6}{ g^6 + c_{w3}^6}\right]^{1/6}
\end{equation}
with 
\begin{equation}
    g = r+c_{w2} \left( r^6 - r\right) \quad  \mathrm{and} \quad 
    r = \frac{\tilde{\nu}}{ \tilde{\mathcal{S}} k^2 d^2}.
\end{equation}
Standard values for the calibration constants $c_{v1}=7.1$ $c_{b1}=0.1355$, $\sigma=2/3$, $c_{b2}=0.622$, $c_{w2}=0.3$, $c_{w3}=2$ are used \citep{spalart1994}, with $c_{w1}=c_{b1}/k^2+(1+c_{b2})/\sigma$ to balance production, diffusion and destruction in the log-layer and with $k=0.41$.

\begin{figure}
    \centering
    \includegraphics[width=\textwidth]{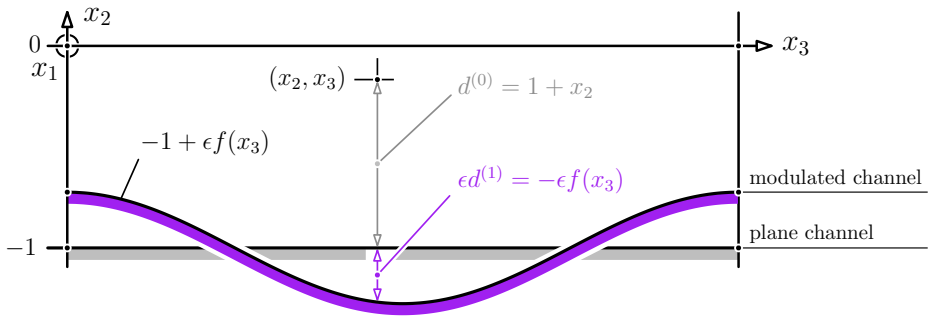}
    \caption{Illustration of the effect of topographic modulations on the distance $d$ appearing in the production and destruction terms of the Spalart-Allmaras transport model. For a point $(x_2, x_3)$ above the trough in the lower channel half, the (positive) distance to the nearest wall increases from $d^{(0)}$, the original distance from the flat lower wall, by an amount $d^{(1)} = -f(x_3)$. Opposite effects are produced on the crests of the topography or in the upper half of the channel.}
    \label{distance_scheme}
\end{figure}

Expanding all flow variables in a Taylor series, the transport equation for the modified eddy viscosity at order zero and one can be obtained. At order zero, the equation is trivially obtained from (\ref{eqn:taulindefinition}) and it is omitted here. At first order, the eddy viscosity $\nu_t^{(1)}$ appearing in the stresses (\ref{tau23qcr}) can be readily obtained as
\begin{equation}
    \nu_t^{(1)}=\tilde{\nu}^{(1)} f_{v1}^{(0)}+\tilde{\nu}^{(0)} f_{v1}^{(1)},
    \label{eq:nut1}
\end{equation}
where $f_{v1}^{(1)}$ and other additional terms appearing at first order are reported in appendix \ref{app:termini}. In the linearisation process, it is key to observe that the topographic modulation can be thought of as a perturbation of the distance from the solid wall. This is a key physical parameter in the SA turbulence model as it controls the formation of a log-layer through the balance of production and destruction, where it appears directly. In particular, the distance is expanded as 
\begin{equation}
    d(x_2, x_3) = d^{(0)}(x_2) + \epsilon d^{(1)}(x_2, x_3),
\end{equation}
with $d^{(0)}$ the original distance in the plane channel and 
\begin{equation}\label{eq:d_1}
d^{(1)}(x_2, x_3) = \mathrm{sign}(x_2) f(x_3),
\end{equation}
where the sign function in \eqref{eq:d_1} captures the symmetric modulation of the walls and models the fact that the distance from the nearest physical wall decreases/increases for points above the crests/troughs of the topography in the lower channel half, as illustrated in figure \ref{distance_scheme}. 

After algebraic operations, the transport equation for the perturbation of the modified eddy viscosity $\tilde{\nu}^{(1)}$ reads as
\begin{eqnarray}
     -\frac{\partial \psi^{(1)}}{\partial x_3} \frac{\partial \tilde{\nu}^{(0)}}{\partial x_2}& =& \frac{1}{\sigma}\left( \frac{1}{R_{e_\tau}}+\tilde{\nu}^{(0)}\right) \left( \frac{\partial^2 }{\partial x_2^2}+\frac{\partial^2}{\partial x_3^2}\right) \tilde{\nu}^{(1)}+\frac{1}{\sigma}\frac{\partial^2 \tilde{\nu}^{(0)}}{\partial x_2^2}\tilde{\nu}^{(1)}  \nonumber\\  &+&\frac{1}{\sigma}(2+2 c_{b2}) \frac{\partial \tilde{\nu}^{(0)}}{\partial x_2} \frac{\partial \tilde{\nu}^{(1)}}{\partial x_2} +c_{b1}\tilde{\nu}^{(0)} \tilde{\mathcal{S}}^{(1)}+c_{b1} \tilde{\nu}^{(1)} \tilde{\mathcal{S}}^{(0)}  \nonumber \\ &-2& \tilde{\nu}^{(0)} c_{w1} f_w^{(0)} \frac{ \tilde{\nu}^{(1)} d^{(0)}- \tilde{\nu}^{(0)}d^{(1)}}{d^{(0)}\,^3}-
    c_{w1} f_w^{(1)} \left( \frac{\tilde{\nu}^{(0)}}{d^{(0)}}\right)^2.
    \label{SAfirstorder1}   
\end{eqnarray}
This equation is coupled to the streamfunction equation by the convective transport term at the left-hand-side, modelling the wall-normal transport of the background turbulent fluctuations by the secondary motions. An additional coupling term with the streamwise momentum equation appears in the production term $\tilde{\mathcal{S}}^{(1)}$, which models the change in the production of turbulent kinetic energy as a result of the distortion of the streamwise velocity profile. 

\subsection{Linearised boundary conditions}
\label{sec:boundarycondition}
Boundary conditions for the linearised transport equations are now derived using established methods \citep{luchini2013,busse2012}. Assuming that the topographic perturbation is small, we retain the original rectangular geometry of the domain but we introduce inhomogenous boundary conditions on the perturbation quantities derived by imposing the original conditions on the displaced surface. 

Considering the lower wall, expanding the velocity near the surface in a Taylor series and enforcing the no-slip condition we obtain
\begin{equation}
    \displaystyle \overline{u}_i(-1+\epsilon f(x_3), x_3) = \left. \overline{u}_i\right|_{x_2=-1}+ \epsilon  f(x_3) \left.\frac{\partial \overline{u}_i}{ \partial x_2}\right|_{x_2=-1} = 0.
    \label{linearbcs}
\end{equation}
Substituting the expansion (\ref{lineardecomposition}) for the velocity in (\ref{linearbcs}), noting that $u_i^{(0)}=0$ at \mbox{$x_2=-1$}, and retaining terms at order one in $\epsilon$ provides
\begin{equation}
      \displaystyle \left. u_{i}^{(1)} \displaystyle \right|_{x_2=-1} +  f(x_3) \left.\frac{\partial u_i^{(0)}}{\partial x_2}\right|_{x_2=- 1}  = 0, \label{u1bcs}
\end{equation}
i.e.~the perturbation velocity at the boundary of the numerical domain is proportional to the wall-normal gradient of the velocity in the plane channel to preserve the no-slip condition on the modulated topography. The boundary condition on the streamwise velocity perturbation then becomes
\begin{equation}
     u_{1}^{(1)}(x_2 = -1) = -  f(x_3) \left.\frac{\partial u^{(0)}}{\partial x_2}\right|_{x_2=- 1} = -  f(x_3) \Rey_{\tau} , \label{eqn:bcsfinal}
\end{equation}
while $u_{3}^{(1)}(x_2=-1) = 0$ and $u_{2}^{(1)}(x_2=-1) = 0$. 
The boundary conditions for the perturbation streamfunction  {are}
\begin{equation}
    \frac{\partial \psi^{(1)}}{\partial x_2}(x_2 = -1) = \psi^{(1)}(x_2 = -1) = 0.
    \label{eqn:bcspsi}
\end{equation}
Using a similar strategy, and noting that the modified eddy viscosity satisfies homogeneous boundary conditions at the wall \citep{spalart1994}, the inhomogeneous boundary condition
\begin{equation}
     \tilde{\nu}^{(1)} (x_2= - 1)=- f(x_3) \left. \frac{\partial \tilde{\nu}^{(0)}}{\partial x_2}\right|_{x_2=-1}=-f(x_3)k
     \label{equn:bcsnut}
\end{equation}
can be derived for the perturbation of the transformed eddy-viscosity at the lower numerical boundary. The last equality holds since the modified eddy viscosity obeys the linear relation $\tilde{\nu} = k x_2$ near the wall \citep{spalart1994}. No conditions are required for the eddy viscosity $\nu_t$, since this is not directly associated to a transport equation in the SA model. With a similar procedure, boundary conditions on the upper numerical boundary can be obtained.  {Equivalently, symmetric boundary conditions can also be applied at the channel centreline when symmetric channels are studied, to reduce computational costs. However, in the present work, the full channel domain with linearised boundary conditions on both upper and bottom surfaces was considered, as it was easily modelled using available Chebyshev discretization tools.}

\subsection{Fourier spectral expansion of the solution} \label{sec:fourier}
When using linearised equations, any arbitrary topography can be analysed by examining each fundamental spanwise length scale separately from the others.
The solution of the linearised equations can be first expressed by the Fourier series
\begin{subeqnarray}\label{eq:expansion}
    u_1^{(1)} (x_2, x_3) &=&\sum_{n=1}^{\infty} \, \hat{u}_1(x_2; n) \cos{(n k_3 x_3)},\\
    \psi^{(1)} (x_2, x_3) &=&\sum_{n=1}^{\infty} \, \hat{\psi}(x_2; n) \sin{(n k_3 x_3)},\\
       \tilde{\nu}^{(1)} (x_2, x_3) &=&\sum_{n=1}^{\infty} \, \hat{\nu}(x_2; n) \cos{(n k_3 x_3)},
\end{subeqnarray}
where $\hat{u}_1(x_2;n)$, $\hat{\psi}(x_2;n)$ and $\hat{\nu}(x_2;n)$ are the real-valued, wall-normal profiles of the perturbation streamwise velocity, streamfunction and modified eddy viscosity at each integer spanwise wavenumber $n$. Then, components at different spanwise wavenumbers decouple, forming the set of three ordinary differential equations
\begin{subeqnarray}
\displaystyle
  -n k_3 \hat{\psi} \Gamma&=& \frac{1}{\Rey_\tau}\left(\frac{\mathrm{d}^2 }{\mathrm{d}x_2^2} -n^2 k_3^2 \right) \hat{u}_1+n k_3 \hat{\tau}_{13}+\frac{\mathrm{d} \hat{\tau}_{12}}{\mathrm{d} x_2}, \\ [16pt]
\displaystyle
 0&=&\frac{1}{\Rey_\tau}\!\left(\!\frac{\mathrm{d}^2 }{\mathrm{d} x_2^2}\!-n^2 k_3^2\!\right)^2\!\! \hat{\psi}-k_3 \frac{\mathrm{d}}{\mathrm{d} x_2}\! \!\left(\!\hat{\tau}_{33}\!-\!\hat{\tau}_{22}\right)\!+\!\left( \frac{\mathrm{d}^2}{\mathrm{d} x_2^2}\!+n^2 k_3^2 \right)\!\hat{\tau}_{23},\\ [16pt]
 -n k_3 \hat{\psi} \displaystyle \frac{\mathrm{d} \tilde{\nu}^{(0)}}{\mathrm{d} x_2}\!&=&\! \frac{1}{\sigma} \!\left(\frac{1}{\Rey_\tau}+\tilde{\nu}^{(0)}\right)\left(\frac{\mathrm{d}^2}{\mathrm{d} x_2^2}\!-\!n^2 k_3^2\right)\hat{\nu} \!+\! \frac{1}{\sigma}\frac{\mathrm{d}^2 \tilde
{\nu}^{(0)}}{\mathrm{d} x_2^2}\hat{\nu} \nonumber \\ &+&\frac{1}{\sigma}(2+2 c_{b2})\frac{\mathrm{d} \tilde
{\nu}^{(0)}}{\mathrm{d} x_2}\frac{\mathrm{d} \hat{\nu}}{\mathrm{d} x_2}+c_{b1} \tilde{\nu}^{(0)} \hat{\tilde{\mathcal{S}}}+c_{b1}\tilde{\mathcal{S}}^{(0)}\hat{\nu} \nonumber\\ &-&2 \tilde{\nu}^{(0)} c_{w1} f_{w}^{(0)} \frac{\hat{\nu} d^{(0)}+\tilde{\nu}^{(0)} f(x_3)}{d^{(0)} \,^3}- c_{w1}f_{w}^{(1)}\left(\frac{\tilde{\nu}^{(0)}}{d^{(0)}}\right), 
 \label{systemequation}
\end{subeqnarray}
along the wall-normal direction at each integer wavenumber $n = 1, 2, \ldots $. In these equations, the wall-normal profiles $\hat{\tau}_{ij}(x_2; n)$ are the components of the Reynolds stress tensor $\tau_{ij}^{(1)}$ obtained by substituting the expansion (\ref{eq:expansion}) into the definitions of the perturbations (\ref{tau23qcr}). This leads ultimately to a set of equations that only contains the quantities $\hat{u}_1(x_2; n)$, $\hat{\psi}(x_2; n)$ and $\hat{\nu}(x_2; n)$. Using the boundary conditions (\ref{eqn:bcsfinal}, \ref{eqn:bcspsi}, \ref{equn:bcsnut}), these variables must satisfy
\begin{subeqnarray}
\label{eq:lin_BC}
    \hat{u}_{1}(x_2=\pm 1) &=& - f^n \, \Rey_\tau,\\
    \hat{\psi} (x_2=\pm 1) &=& \mathrm{d}\hat{\psi}/\mathrm{d}x_2 (x_2=\pm 1) = 0,\\
    \hat{\nu} (x_2=\pm 1) &=& - f^n k.
\end{subeqnarray}

Inspection of these boundary conditions and the governing equation shows that the wall topography affects the formation of secondary flows with three separate forcing terms. The first mechanism is mediated by the distance perturbation $d^{(1)}=-f(x_3)$. This term appears directly in the linearised transport equation of the eddy viscosity as a source term, suggesting that the topography modulation is felt throughout the domain as an alteration of the wall-normal development of the turbulent stresses. Crucially, spanwise heterogeneity of the topography produces a spanwise modulation of the eddy viscosity, i.e. of the Reynolds stress, which is known to be a source term in the transport equation of the turbulent kinetic energy \citep{barros2014, Hwanglee2018}. The second and third mechanisms are localised at the wall and are controlled by the inhomogeneous boundary conditions on the streamwise velocity and the perturbation eddy viscosity, respectively. The former produces a positive/negative velocity slip on the trough/crests of the modulation and generates a streaky motion with the associated streamwise velocity spanwise gradients. All these forcing terms are proportional to the strength of the coefficient $f^n$ in the series \eqref{fx3} characterising the surface geometry, showing the importance of fully characterising the spectral content of the wall topography.

The numerical solution of the system (\ref{systemequation}) with the boundary conditions (\ref{eq:lin_BC}) is obtained by discretising the equations over \mbox{$x_2 \in [-1, 1]$} using a Chebyshev-collocation method. A spectral technique is technically not ideal for this problem, because $d^{(0)}$ has a sharp cusp at $x_2=0$. Nevertheless, we have observed that the spectral technique is robust in practice and provides accurate results when a sufficiently fine collocation grid is utilised. In the following calculations, we used no less that 202 collocation points, progressively increasing the resolution at the higher Reynolds numbers considered. The numerical code was also validated on sinusoidal channels using a nonlinear SA-QCR custom implementation in OpenFoam, with good agreement.

\subsection{Reynolds-averaged solution in plane channels} \label{meanflowstructure}
The profiles of the mean streamwise velocity and the eddy viscosity of the plane channel appear in the first-order equations (\ref{systemequation}) and are thus shown in this section. Profiles of these quantities were obtained by solving the SA equation (\ref{SAequation1}) coupled with the streamwise momentum equation (\ref{RANS}) on a one-dimensional domain extending in the wall-normal direction using an in-house code. A linear Boussinesq approach is used, as this is sufficient in plane channels. The numerical code is based on a Chebyshev-collocation discretization and uses a Jacobian-free Newton–Krylov technique to solve the nonlinear coupled system of algebraic equations \citep{knollkeyes2004}.

\begin{figure}
    \centerline{
    \includegraphics[width=0.88\textwidth]{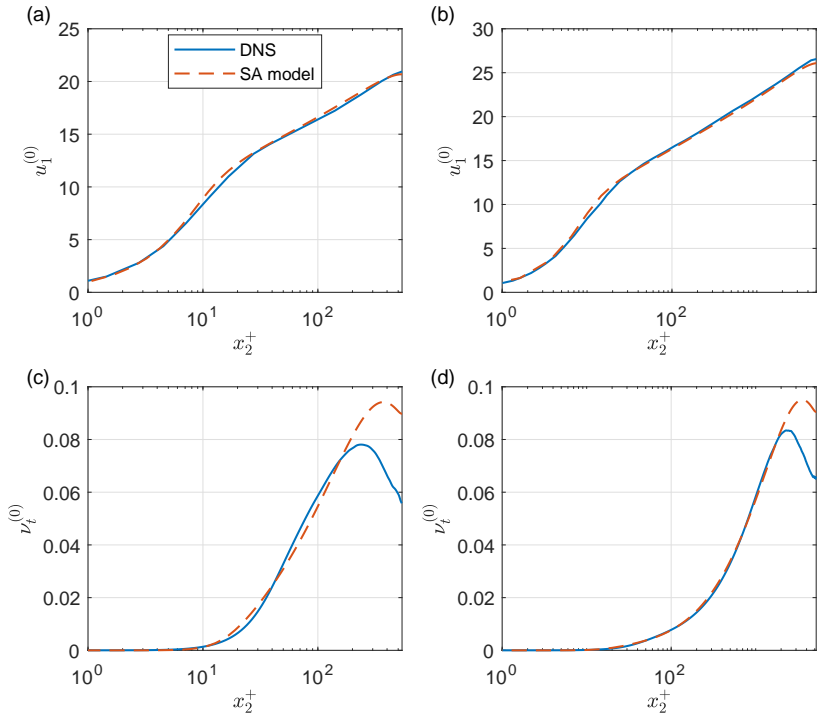}}
    \caption{Profiles of streamwise velocity, top panels, and of the turbulent eddy viscosity, bottom panels, in plane channel from the SA model ($----$) and from the DNS (----) of \citet{leemoser2015}. Data is shown for $\Rey_{\tau}=550$ in panels (a, c) and $\Rey_{\tau}=5200$ in panels (b, d). }
    \label{zeroordercomparison}
\end{figure}

Mean streamwise velocity profiles obtained from the RANS solver at \mbox{$\Rey_{\tau}=550$} and \mbox{$\Rey_{\tau}=5200$} are shown in figure \ref{zeroordercomparison}(a) and \ref{zeroordercomparison}(b), respectively, as a function of the wall normal distance $x_2^+$ scaled by the viscous length (dashed red lines). These profiles extend to the channel mid-plane and are compared with the direct numerical simulation results of \citet{leemoser2015} (solid blue lines). The SA solution agrees well with the DNS data, especially in the logarithmic layer, although higher velocities are observed in the buffer layer region. Profiles of the turbulent eddy viscosity $\nu_t^{(0)}$ are shown in figure \ref{zeroordercomparison}(c) and \ref{zeroordercomparison}(d), for the same Reynolds numbers. The eddy viscosity is extrapolated from the DNS simulation data by dividing the turbulent stress $- \overline{u_1'u_2'}$ with the wall-normal gradient of the streamwise velocity $\Gamma$. Good agreement with the DNS data is observed, although larger deviations are observed for $|x_2| \gtrsim 0.4$.

\section{Secondary flows in sinusoidal channels} \label{sinusoidalwall}
Secondary flows in symmetric channels with sinusoidal walls (see figure \ref{sketch}(a)) are now considered to elucidate the fundamental role of the spanwise length scale on the generation of secondary flows. This insight can then be used to analyse surfaces with complex spatial characteristics \citep{anderson2015,barros2014}. We consider modulations expressed by the cosine law
\begin{equation}\label{eqn:sinusoidal}
 f(x_3) = \lambda_3  \cos(k_3 x_3).
\end{equation}
Scaling the amplitude with the period $\lambda_3$ ensures that the aspect ratio of the modulation (peak-to-peak amplitude to spanwise length scale) remains constant, i.e., we follow the shallow-roughness limit introduced in \citet{luchini2013}.

\subsection{Organization of secondary currents}\label{sec:organization_of_secondary_currents}
\begin{figure}
    \centerline{
    \includegraphics[scale=0.65]{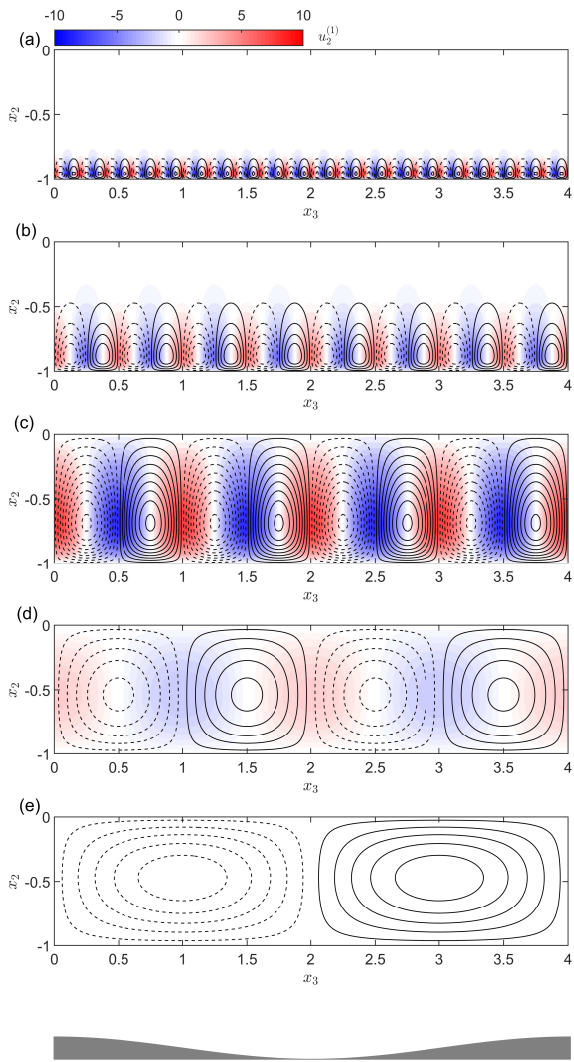}}
    \caption{ {Contours of the perturbation streamfunction $\psi^{(1)}$ 
    in the cross-plane $(x_2, x_3)$ at $\Rey_\tau=5200$ and varying wavelength:  {panel (a) $\lambda_3=0.2$, panel (b)  $\lambda_3=0.5$, panel (c) $\lambda_3=1$, panel (d) $\lambda_3=2$ and panel (e) $\lambda_3=4$}. The streamfunction perturbation is limited to [-1, 1] in panels (a) and (b), to [-2, 2] in panels (c) and (d), to [-0.2, 0.2] in panel (e) for a better representation of the flow structures. Dashed lines are used for negative values.} The colour map of the wall-normal velocity perturbation (in units of the friction velocity and per unit of modulation amplitude) is also reported.  {For wavelengths smaller than $\lambda_3=4$, the flow topology for a single period is repeated to better display the evolution in size and strength of secondary flows. The topography from crest-to-crest is illustrated for the sake of clarity for $\lambda_3=4$ in the bottom.} }
    \label{fig:topology}
\end{figure}

The flow topology predicted by the linearised model is visualised in figure \ref{fig:topology} for $\lambda_3 = 0.2, 0.5, 1, 2$ and $4$, in panels (a) to (d), respectively. Contours of the perturbation streamfunction (dashed contours for negative values) are reported. The colour map shows the wall-normal component $u_2^{(1)}$. Data at a large Reynolds number, $\Rey_\tau = 5200$, is reported as an illustrative example. Reynolds number effects are discussed later. A sketch of the harmonic topography is also reported at the bottom for $\lambda_3 = 4$. For the symmetric configuration considered here, only data in the lower half of the channel is shown. 
The predicted secondary structure displays two counter-rotating vortices per period in the lower half of the channel. 
 {A similar flow organization was recently observed by \citet{vidal2018} using direct numerical simulations on wavy channels with antisymmetric walls. However, the present results refer to a symmetric channel where the vortices are confined to the half channel height. On the contrary, in the simulations carried out by \citet{vidal2018} the vortices extend from the bottom to the upper wall. In addition, finite modulation amplitudes, with non-negligible convective effects, are considered \citet{vidal2018}, unlike in the present case, where the modulation is infinitesimal. Nevertheless, in both cases the vortices flank the crest of the modulation and produce an upwelling motion above the crests.}
Conservation of mass through the channel then implies that a downwash is observed in the troughs of the topography. The height of the region affected by the secondary motion increases with $\lambda_3$ and, eventually, the vortices occupy the full half-height of the channel for $\lambda_3 \approx 1$. This topology persists from low periods up to $\lambda_3 \approx 6$, beyond which a large-scale flow reversal,  {where secondary currents rotate in the opposite direction and produce a downwash over the crest}, is observed.  {This phenomenon is not related to the appearance of tertiary flows \citep{vanderwel2015, Hwanglee2018, medjnoun_vanderwel_ganapathisubramani_2020} and might be a product of the turbulence model utilised in this paper that would not be observed in DNS or experiments. However, data to validate or disprove this behaviour for modulations with such large period does not seem to be available in the literature and further investigation is warranted.} 

\subsection{Velocity profiles}
\begin{figure}
    \centering
    \includegraphics[width=\textwidth]{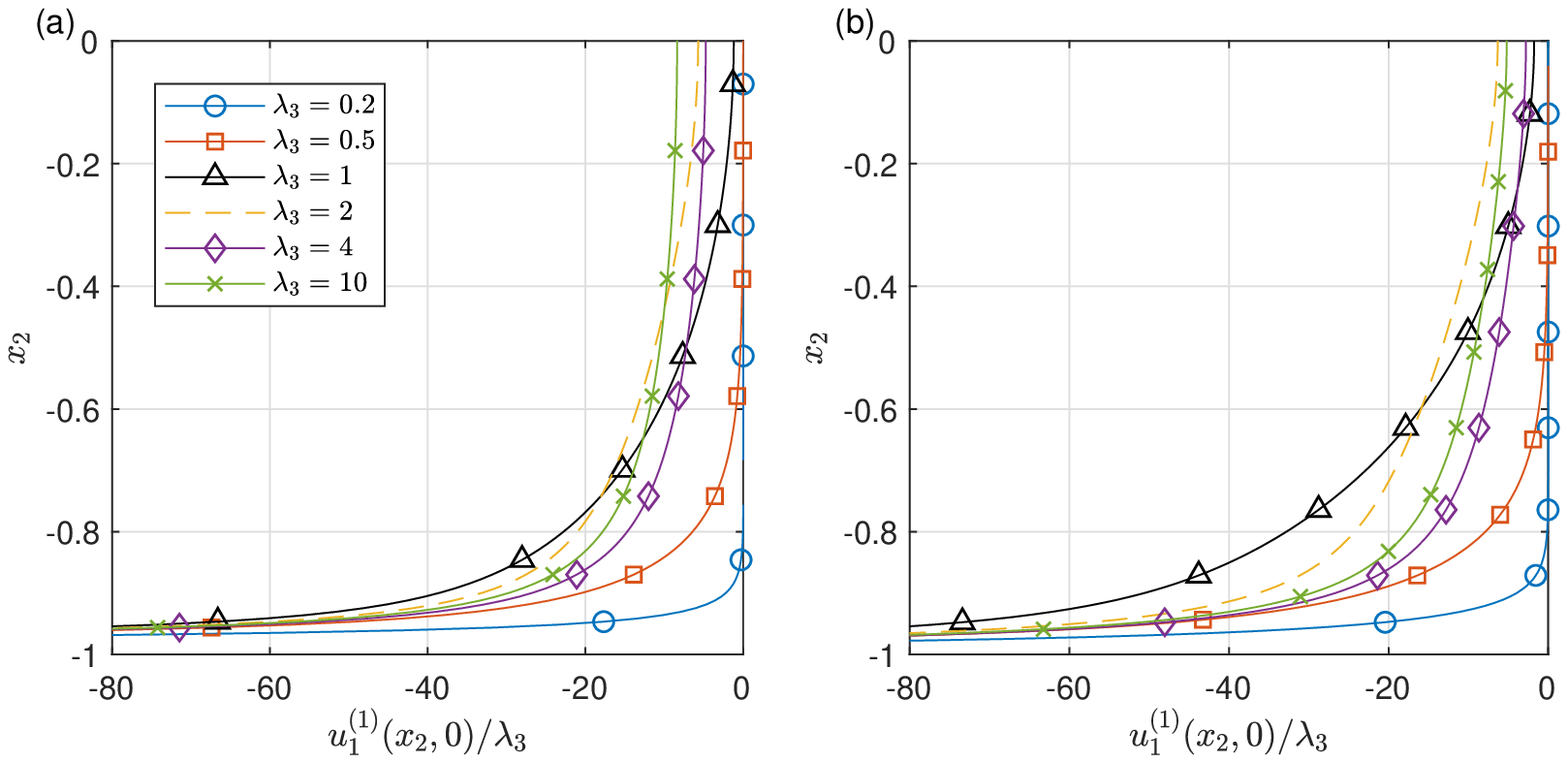}
    \caption{Streamwise velocity perturbation per unit of wall-modulation amplitude $u_1^{(1)}(x_2, 0)/\lambda_3$ at $Re_\tau=550$ (panel a) and $Re_\tau=5200$ (panel b) from $\lambda_3=0.2$ to $\lambda_3=10$. The velocity profiles are extracted above the modulation crest. The velocity axis is restricted to $[-80, 0]$ for clarity, since the velocity perturbation at the lower domain boundary is $-\Rey_\tau$.}
    \label{fig:ciccio}
\end{figure}
 {Wall-normal profiles of the streamwise velocity component for $\lambda_3=0.2, 0.5, 1, 2, 4$ and $10$ are reported in figure \ref{fig:ciccio} at  $\Rey_\tau=550$ in panel (a) and at $\Rey_\tau=5200$ in panel (b). These profiles are localised at $x_3=0$, on the crest of the modulation. Velocity profiles at any other spanwise location, e.g. over the trough, can be obtained by utilising the expansion (\ref{eq:expansion}) restricted to a single spanwise wavenumber mode.  Given that the amplitude of the wall modulation \eqref{eqn:sinusoidal} is proportional to $\lambda_3$, the velocity is first scaled by the wavelength and it should thus be interpreted as the flow response per unit amplitude of modulation expressed in terms of $h$. The velocity perturbation at the lower domain boundary is equal to $-\Rey_\tau$ due to the boundary conditions \eqref{eqn:bcsfinal}. We observe that the streamwise velocity is always negative, for all periods considered, corresponding to a low momentum pathway (LMP) over the crest, as previously observed by other authors, e.g. \citet{vanderwel2015} among others. The velocity perturbation decreases in magnitude when moving towards the channel centre. The depth of the disturbance increases with $\lambda_3$, as the secondary structures grow in size. The effect of the Reynolds number is moderate and consists of a slight increase in the momentum deficit when comparing corresponding profiles in panels (a) and (b).}

\begin{figure}
 \centering
    \includegraphics[width=\textwidth]{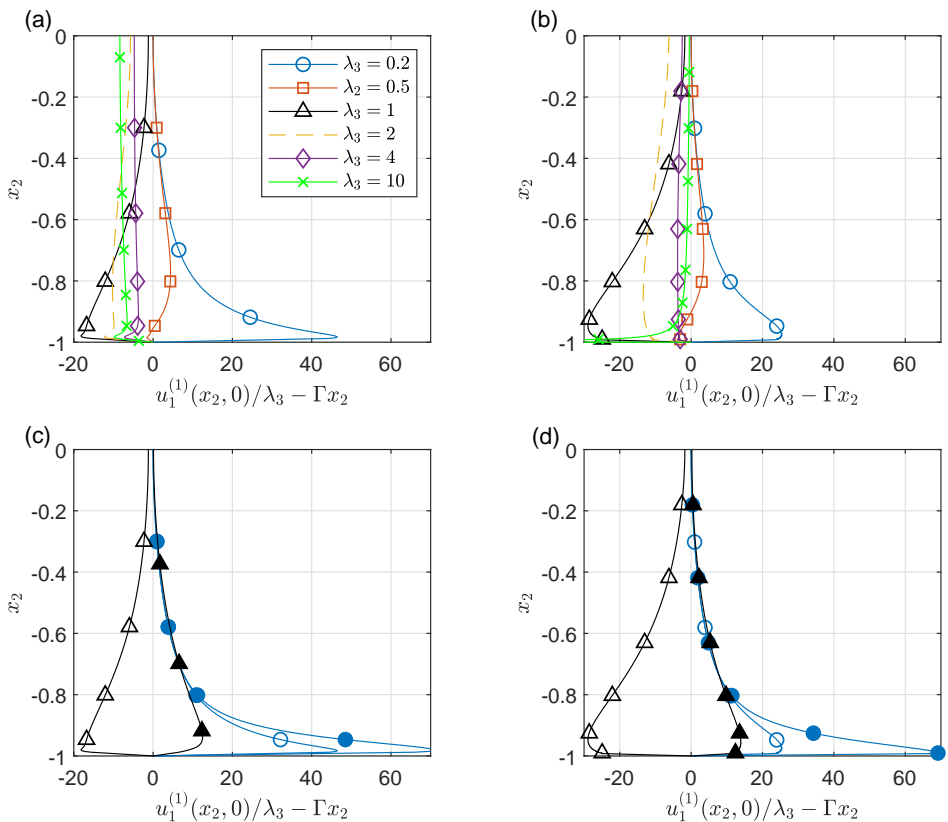}
    \caption{Profiles of  {the modified  streamwise velocity perturbation}  $u_1^{(1)}(x_2, 0)/\lambda_3 - \Gamma x_2$ at $\Rey_\tau=550$, panel (a), and $\Rey_\tau=5200$, panel (b), at different spanwise wavelengths. Profiles are located over the modulation crest. In the figure, ---$\circ\!\!\!$--- $\lambda_3=0.2$, ---$\square\!\!\!\!$--- $\lambda_3=0.5$, ---$\!\!\!\bigtriangleup\!\!\!$--- $\lambda_3=1$, $---$ $\lambda_3=2$, ---$\!\!\Diamond\!\!\!$--- $\lambda_3=4$  { and  ---$\!\!X\!\!$--- $\lambda_3=10$}. In panels (c, d), the effect of turning on/off the QCR strain-stress model is shown for the same Reynolds numbers. Symbols are the same as in panel (a) but filled symbols are used for solutions at $C_{r1}=0$.}
    \label{fig:streamwise_velocity}
\end{figure}

 {To better elucidate how the wall modulation alters the spatial structure of the streamwise velocity component, the quantity $\Gamma x_2$ is subtracted from the profiles of figure \ref{fig:ciccio}. This quantity attempts to capture the velocity perturbation produced by the shift in the mean velocity profile when the wall is displaced, particularly strong in the near-wall region but formally zero at the mid-plane. Results are reported in panels (a) and (b) of figure \ref{fig:streamwise_velocity}.}
It can be observed that the streamwise velocity perturbation is more pronounced in the near-wall region and relatively less in the channel centre. For short periods, this perturbation is positive, indicating that the near-wall flow over the crests moves faster than it would do over a flat wall.
By contrast, for larger periods, the streamwise velocity perturbation is negative, initially in the vicinity of the wall and then gradually across the full channel half-width.

The change of sign with $\lambda_3$ suggests that two competing mechanisms are at play. The first mechanism is originated from the vertical ``protrusion'' of the crests towards the mid plane, causing higher velocity over the crests ``exposed'' to the bulk of the flow. The second mechanism is the up-welling/down-welling motion introduced by the secondary structures. As shown in figure \ref{fig:topology}, these structures transport low momentum fluid from the near-wall region over the crest  upwards toward the channel core, causing a local reduction of the flow velocity and vice versa over the troughs. When $\lambda_3$ is sufficiently large so that secondary currents are strong enough and they span a sufficiently large fraction of the channel, this second effect prevails and a low speed streak forms over the crests between the streamwise rolls, similarly to the optimal roll/streak configuration found in shear flows \citep{delalamo2006, pujals2009}. 

To better quantify the strength of these two competing mechanisms, we report in panels (c) and (d) of figure \ref{fig:streamwise_velocity} the streamwise velocity profiles obtained from calculations where the QCR constant $C_{r1}$ is set to zero (filled symbols), corresponding to using a linear Boussinesq's stress/strain relation. From a practical perspective, this is equivalent to ``turning off'' secondary motions, so that only the first mechanism is active. The profiles are compared to the reference case at $C_{r1}=0.3$ (open symbols) and data for $\lambda_3=0.2, 1$ at the same Reynolds numbers of panels (a) and (b) is shown. When $C_{r1} = 0$, the velocity perturbation is always positive due to the protrusion of the crests into the bulk of the flow, as just mentioned, but when the QCR model is activated, negative velocities can be observed.

A further remark is that the profiles of the streamwise velocity show that the local wall shear stress perturbation can be significant. However, the perturbation of the spanwise-averaged shear stress predicted by the present linearised model is identically zero. In fact, from the expansions (\ref{eq:expansion}), it is easy to show that the wall shear stress is simply a harmonic function, with zero mean. A linear method cannot predict changes in spatially-averaged quantities for flows obeying  {translational} symmetries as in the present case, and second order effects (i.e. large perturbations) must be taken into account to uncover, e.g., how drag is affected by topography changes. However, having zero-mean velocities does not imply that spatial averaging is trivial for all other quantities. For instance, the spanwise averaged dispersive stresses often reported to characterise the secondary currents \citep{smith1977,raupach1982,nikora2001} can be non-zero. 
 {It is worth noting that the effective streamwise velocity profile resulting from the wall modulation is given, in our framework, by the sum of the flat channel profile and a small perturbation produced by the wall modulation. The streamwise velocity perturbation reported in figure \ref{fig:streamwise_velocity} does not show any logarithmic behaviour, as it is the product of the two competing mechanisms discussed above. This might explain the strong distortion of the log-layer behaviour often observed in experiments or simulations of flows over heterogeneous surfaces \citep{medjnoun_vanderwel_ganapathisubramani_2018}.}

\begin{figure}
    \centerline{
    \includegraphics[width=\textwidth]{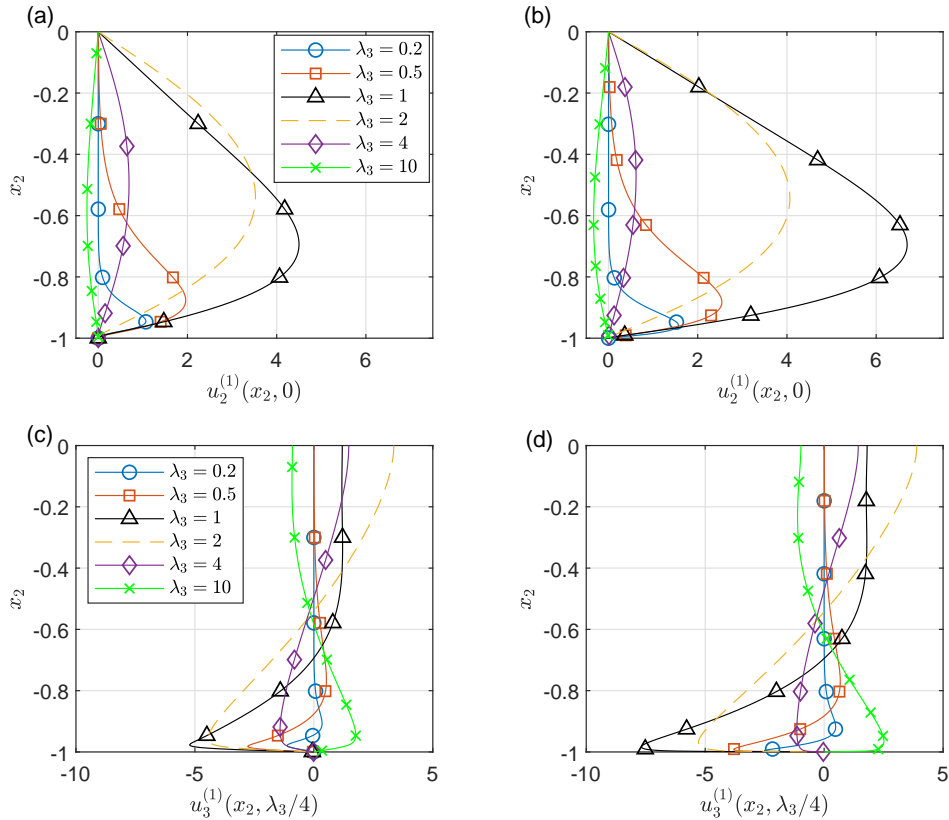}}
    \caption{Comparison of the profiles of the velocity components for different $\lambda_3$, at $\Rey_\tau=550$ in panels (a,c) and $\Rey_\tau=5200$ in panels (b,d).  In panels (a,b) the wall-normal velocity $u_2^{(1)}(x_2,0)$ is plotted, in (c,d) the spanwise velocity $u_3^{(1)}(x_2,\lambda_3/4)$ is plotted.  In the figure, ---$\circ\!\!\!\!$--- $\lambda_3=0.2$, ---$\!\!\square\!\!\!$--- $\lambda_3=0.5$, ---$\!\!\!\bigtriangleup\!\!\!$--- $\lambda_3=1$, $---$ $\lambda_3=2$, { ---$\Diamond\!\!\!$--- $\lambda_3=4$ and ---$\!\!X\!\!$--- $\lambda_3=10$.}} 
    \label{fig:vel_components}
\end{figure}

Profiles of the wall-normal and spanwise velocity components at $x_3=0$, on the crest of the modulation, and $x_3=\lambda_3/4$ respectively, are reported in figure \ref{fig:vel_components} for the same Reynolds numbers and wavelengths considered in figure \ref{fig:streamwise_velocity}. As anticipated, in the lower half of the domain, the linearised RANS model predicts positive wall-normal velocities, indicating an upwash on the crest of the modulation and a downwash in the trough produced by secondary currents induced by the topography. For short periods, these effects are confined near the wall but the depth of the region influenced by this upwelling motion increases with the spanwise period up until $\lambda_3 \approx 1$, where the wall-normal motion involves the entire channel half-height. When the spanwise length scale is further increased, the wall-normal velocity decreases, as the vortical structures do not have additional space to grow. 

 {For $\lambda_3=10$, the direction of rotation of the secondary vortices, occupying the entire channel half-height, is opposite to what is observed at the lower periods shown in figure \ref{fig:topology} and downwash is now observed over the crest. Interestingly, the streamwise velocity profile in figure \ref{fig:ciccio} is still negative, i.e. the flow displays a LMP associated to a downwash. Although this might appear to contradict established knowledge, the downwash velocity is relatively small in magnitude. It is argued that the negative streamwise velocity is purely a result of the contraction of the channel height over the crest (symmetric channels are considered), since the wall-normal velocity is too weak in this case to justify the observed change in the streamwise velocity. This effect would not be observed in an open flow like a boundary layer, where the wall modulation does not produce a contraction of the available cross-sectional area.} 

For the spanwise velocity, strong negative values are observed near the wall on the right flank of the harmonic topography, producing a lateral jet-like motion towards the modulation crest. Generally, the negative velocity peak is larger than the peak of positive velocity, due to the confinement of the vortices near the wall  {(see panels (c) and (d) in figure \ref{fig:vel_components}).} The peak location varies only modestly with $\lambda_3$, but it gets closer to the wall and more intense at larger Reynolds numbers.  {For $\lambda_3=10$, the spanwise velocity profile shows a positive peak in the near wall region due to the change of direction of rotation discussed previously. }

\subsection{Effect of wavelength and Reynolds number on the intensity of secondary flows} \label{sec:gainplot}
\begin{figure}
    \centerline{
         \includegraphics[width=\textwidth]{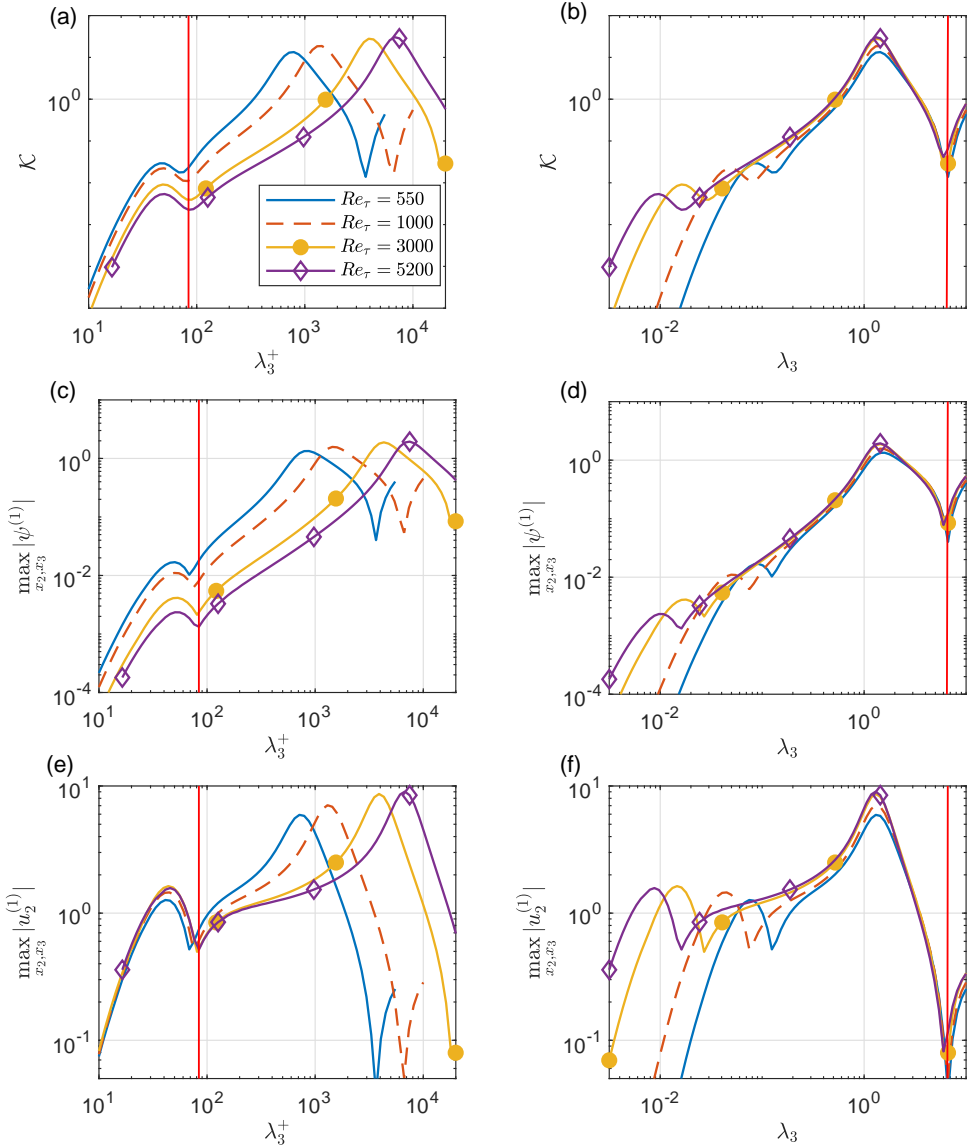}}
    \caption{Intensity of secondary flows as a function of the spanwise wavelength. Different intensity metrics are compared. The panels (a,b) display  the kinetic energy density $\mathcal{K}$, panels (c,d) the maximum streamfunction $\max_{x_2,x_3} | \psi^{(1)}|$ and panels (e,f) the maximum of the wall-normal velocity $\max_{x_2,x_3} | u_2^{(1)}|$. The wavelength is scaled in inner units in (a,c,e) and outer units in (b,d,f). In figure, ------  $\Rey_\tau=550$, $-----$ $\Rey_\tau=1000$,
    ---$\bullet\!\!\!\!$--- $\Rey_\tau=3000$ and ---$\Diamond$--- $\Rey_\tau=5200$. The vertical lines denote particular spanwise length scales where a change in the flow structure (flow reversal) is predicted.}
    \label{kinetic_sinusoidal}
\end{figure}

We now turn to investigating in more depth the effect of the wavelength $\lambda_3$ and of the Reynolds number on the strength of the secondary flows. For this purpose, we utilize the volume averaged kinetic energy of the cross-flow components 
\begin{equation}
\mathcal{K} = \frac{1}{4\lambda_3}\! \int_{-1}^{1} \int_{0}^{\lambda_3} \left[ u_2^{(1)}(x_2, x_3)^2 + u_3^{(1)}(x_2, x_3)^2\right ] \,\mathrm{d}x_3 \,\mathrm{d}x_2,
\label{eq:kappa}
\end{equation}
to characterize the global amplitude of secondary flows. We also utilize the peak value of the perturbation streamfunction $\max_{x_2, x_3} |\psi^{(1)}(x_2, x_3)|$, following \citet{vidal2018}, to characterise the flow rate associated to the vortical flow and the peak wall-normal velocity $\max_{x_2, x_3} |u^{(1)}_2(x_2, x_3)|$. Results are reported in figure \ref{kinetic_sinusoidal}. In the left panels, the dimensional spanwise period is scaled with the viscous length, i.e. $\lambda^+_3 = \lambda_3 \Rey_\tau$, while in the right panels the dimensional spanwise period is scaled with the outer scale $h$. Data for several Reynolds numbers, spanning the range $\Rey_\tau = 550$ to 5200 is reported. The vertical red lines denote regions where the predicted qualitative behaviour changes and are discussed later on.
The key result is that the linearised model predicts two amplification peaks, indicating that the response of the turbulent wall-bounded flow to a harmonic topography modulation is stronger at preferential spanwise length scales. The location of these peaks is weakly dependent on the metric employed. In particular, the location of the first peak collapses when the wavelength is expressed in outer units to a value of $\lambda_3 \approx 1.54$. This peak is associated to large-scale vortical structures that occupy the entire half-height of the channel and produce a significant wall-normal transport through intense upwash/downwash regions on the crests/troughs of the modulation, as described in figure \ref{fig:topology}.  On the other hand, the location of the second peak collapses when scaled in inner units, at $\lambda_3^+ \approx 45$.  We have tested that the constant $C_{r1}$ of the QCR model does not affect the location of these peaks, but only their amplitude. 

This behaviour mirrors the predictions of transient growth analysis reported by \citet{delalamo2006} and \citet{pujals2009} for plane channels. However, the location of the inner peak predicted in the present case is
approximately half of the value found from the transient analysis, i.e. $\lambda_3^+ \approx 100$, which is predictive of the spanwise spacing of near-wall velocity streaks. It is also lower than what proposed by the conceptual model of  \citet{vidal2018} who suggested an inner peak at $\lambda_3^+\approx 130$. 

\begin{figure}
    \centerline{
         \includegraphics[width=\textwidth]{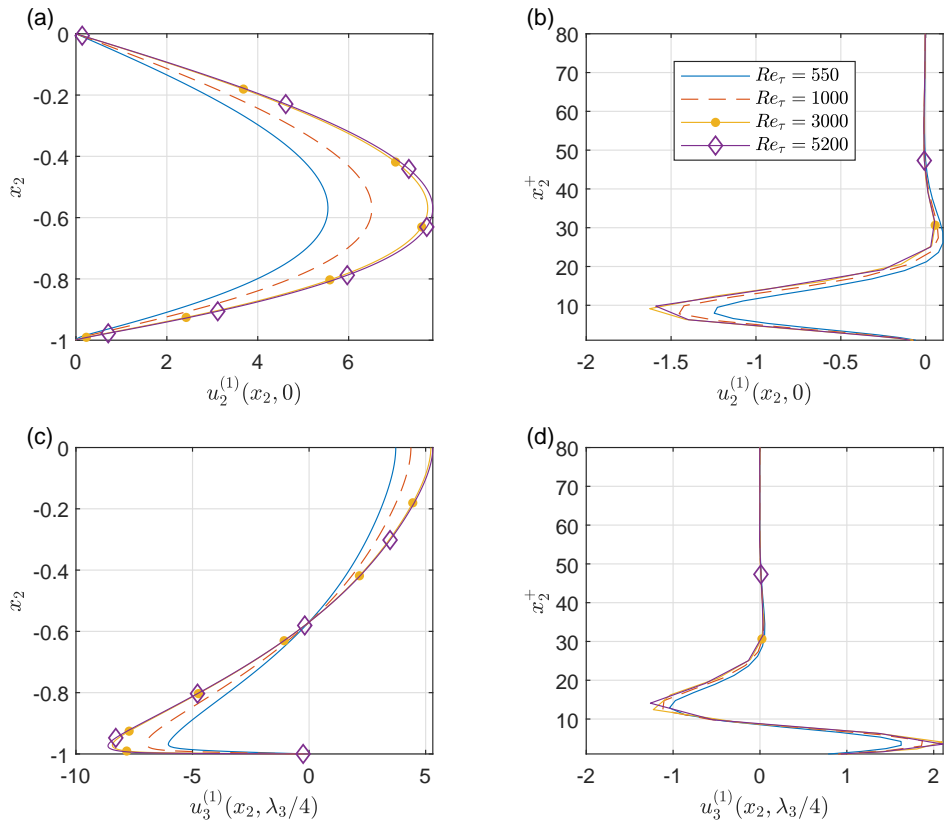}}
    \caption{Wall-normal (top panels) and spanwise (bottom panels) velocity profiles for the outer peak at $\lambda_3=1.54$ in (a, c) and inner peak at $\lambda_3^+=46.5$ in (b, d) for increasing Reynolds number. In the figure, ------  $\Rey_\tau=550$, $-\cdot-\cdot-$ $\Rey_\tau=1000$,
    ---$\bullet\!\!$--- $\Rey_\tau=3000$ and  ---$\Diamond\!\!$--- $\Rey_\tau=5200$. }
    \label{peakscaling}
\end{figure}

To further characterise these amplification peaks, profiles of the wall-normal and spanwise velocity components are reported in figure \ref{peakscaling} for the outer peak (left panels) and inner peak (right panels). The Reynolds number varies from $550$ to $5200$. There are two major observations. Firstly, the present model predicts that the flow response to the surface modulation becomes, asymptotically, independent of the Reynolds number when scaled with the friction velocity, for both the inner and outer peaks. This is a major difference from transient growth analysis, where the energy gain increases with the Reynolds number. More importantly, this result is also in contrast with the findings of \citet{vidal2018} (and references therein) who performed direct numerical simulations in wavy channels and showed that secondary flow velocities scaled by the bulk velocity are not sensitive to the Reynolds number if the Reynolds number is large enough to prevent marginally turbulent flow effects. The predictions of the present model can be attributed to fundamental properties of the SA model used in this study, as already indicated by \citet{spalart2018}. In fact, the SA model is built in order to obtain a collapse of the eddy viscosity profile in the logarithmic layer, where the transport equation (\ref{SAequation1}) has solution $\Tilde{\nu} = k x_2$, as well as in the outer layer. This implies that the eddy viscosity profile, and thus the Reynolds stresses driving the formation of secondary flows of equation (\ref{tau23qcr}) are also, asymptotically, Reynolds number independent when scaled with the friction velocity. 

The second major observation is that the flow topology predicted by our model for the inner peaks is characterised by a downwash over the crest of the modulation, confined in the near-wall region ($x_2^+ < 30$). In fact,  all quantities shown in figure \ref{kinetic_sinusoidal} display two low amplification regions: one at $\lambda_3^+\approx 10^{2}$ and one at $\lambda_3\approx 6$, as denoted by the vertical lines in figure \ref{kinetic_sinusoidal}. At these spanwise length scales, a structural change in the topology predicted by the present model is observed, where a downwash is predicted over the crests of the modulation for either very large or very small wavelengths. While data for very large wavelengths, $\lambda_3 > 6$, does not appear to be presently available in the literature to compare our model with, the flow past surface corrugations at $\lambda_3^+ \approx 50$ is well know (e.g. \citet{goldstein1998,choi1993,chu1993}) and an upwash is typically observed over the crests of the corrugations. The origin of this discrepancy and of the difference in the location of the inner peak compared to what found from transient growth analysis, can be attributed to the fact that the present RANS-based model is likely not able to capture correctly the nature of the interaction between the surface modulations and near-wall turbulent structures when these have commensurate lengths.

 \subsection{Large-scale motions in wall-bounded shear flows and secondary currents.}
The secondary currents predicted by the present model (figure \ref{fig:topology}) and their amplification as a function of the spanwise length scale (figure \ref{kinetic_sinusoidal}) are reminiscent of the optimal structures found with transient growth analysis in flat-wall turbulent channels by various authors \citep{delalamo2006,pujals2009}. These smooth-wall analyses have demonstrated that the Navier-Stokes operator linearised around the turbulent mean profile and augmented with an eddy viscosity term can support transient energy amplification at two specific spanwise length scales, scaling in inner and outer units, respectively. Specifically, streamwise-elongated roll-like motions introduced as initial conditions of the initial value problem develop into longitudinal streamwise streaks. These analyses have provided a formal description of the ubiquitous presence of near-wall streaky motions and  large-scale structures in the outer layer of turbulent shear flows. The underlying mechanism is well-known, i.e., the constructive interaction of nearly-parallel stable eigenfunctions of the Orr-Sommerfeld-Squire equations \citep{butler1993}.  It was recently proposed by \citet{chung2018} that a lateral variation of surface attributes may act a ``phase lock'' to hold naturally-occurring large-scale, outer layer structures around a fixed spatial location. Our linear operator analysis clarifies the relation between these structures and secondary currents, following the view expressed in, e.g., \citet{adrian2012}. Spanwise heterogeneity of surface attributes may be interpreted as a \emph{steady} forcing on the linearised operator, which then produces strong secondary motions with non-zero time average, locked on to the surface modulation. On the contrary, large-scale, outer-layer structures may be interpreted as the manifestation of the amplified response to \emph{unsteady} turbulent velocity fluctuations, as proposed by \citet{delalamo2006, pujals2009}, without spatial locking and hence with zero time-average. Hence, both types of structures may be interpreted as the independent manifestation of the amplification properties of the same linear operator, subjected to either steady or unsteady perturbations.

\section{Secondary flows above rectangular ridges}\label{ridgeswall}
Secondary flows above rectangular ridges are now considered. As shown in figure \ref{sketch}(b), the geometrical parameters considered are the spacing between the ridges $S$ and the ridge width $W$. The gap between the elements is $G=S-W$. Linearised flow solutions in this geometry are obtained wavenumber-by-wavenumber as discussed in section \ref{sec:fourier}. Except for very near the wall, the solution is smooth and the Fourier expansion \eqref{eq:expansion} converges rapidly. To improve the convergence of our spectral code in the near wall region, the discontinuous wall geometry is approximated by the smooth function
\begin{equation}\label{eq:ridge_geometry}
f(x_3)=\frac{1}{\arctan(\alpha)} \arctan\left( \alpha \left[ \cos(k_3 x)-\cos \left (k_3 \frac{W}{2} \right) \right]\right),
\end{equation} 
where $\alpha$ is used to round the corners of the ridges and to increase the roll-off of the coefficients $f^n$ of its cosine series \eqref{fx3}. Here, $\alpha$ is chosen so that $\mathrm{d}f / \mathrm{d}x_3(W/2) = 2\,\cdot 10^4$. The surface geometry is then discretised with at least $150$ cosine waves, ensuring that the ratio $|f^1/f^{150}|$ is no less than 300. We have repeated some calculations at finer resolutions, and no appreciable change in the structure of large-scale motions developing over this geometry has been observed.

\subsection{Effect of geometrical parameters}
\begin{figure}
    \centerline{
    \includegraphics[width=1.05\textwidth]{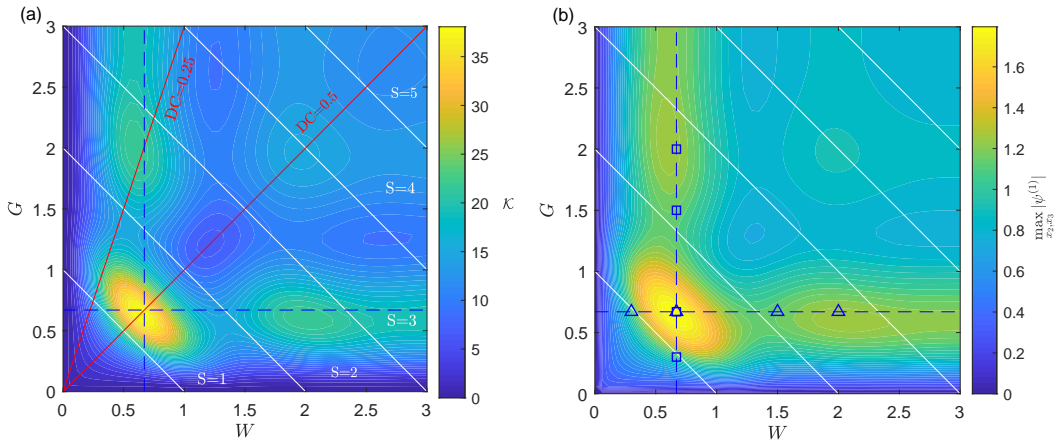}}   
    \caption{Contours of the volume averaged kinetic energy of the cross-stream plane velocities $\mathcal{K}$ (panel a) and streamfunction peak value $\max_{x_2,x_3}|\psi^{(1)}|$ (panel b)  as a function of the gap $G$ and ridge width $W$. The Reynolds number is $\Rey_\tau=5200$. 
    In panel (a), cases at constant duty cycle $DC= 0.25$ and $0.5$ are identified by the red lines. Cases at constant spacing $S=1, 2, 3, \ldots$ are identified by the white lines. Dashed lines identify cases at constant gap or width, with markers for configurations discussed later in the text.}
    \label{map_re}
\end{figure}
To elucidate the role of relevant parameters, we use the kinetic energy density of the cross-stream components, defined in equation (\ref{eq:kappa}), to characterize the global response in the cross-stream plane, and the peak streamfunction value $\max_{x_2,x_3}| \psi^{(1)}(x_2, x_3)|$ to characterize the flow rate associated with the cross-stream motions \citep{vidal2018}.
These two quantities are reported in figure \ref{map_re} as a function of the width $W$ and the gap $G$. Configurations at constant spacing $S = G + W = 1, 2, 3, \ldots $ lie on the white lines with slope $-1$. Note that configurations at constant duty cycle  {$DC = W/S$}, considered as a relevant parameter in e.g. \citet{castro2020}, lie on straight lines passing through the origin with slope  {$1/DC-1$}. 
Results for $\Rey_\tau=5200$ are reported, since, as discussed in section \ref{sec:gainplot}, the SA-based RANS model predictions are asymptotically Reynolds number independent, and no qualitative changes to the following discussion arise when the response at other Reynolds numbers is examined.

Regardless of the metric used, secondary motions are weak for $S < 1$ and their strength peaks at $S \approx 1.34$, close to that obtained for sinusoidal walls and in agreement with predictions obtained in experiments on rectangular ridges \citep{medjnoun_vanderwel_ganapathisubramani_2020} but also for secondary flows developing over roughness strips \citep{chung2018, wangsawijaya2020} and streamwise arrays of roughness elements \citep{yang2017}. 
The contours of the response have a preferential orientation whereby weaker changes in the response are observed when the spacing $S$ is held constant at the optimal value and $W$ and $G$ are varied. This occurs because such surfaces have a strong periodic component at the optimal length scale $S\approx1.34$. This explains why many studies have identified this length scale as producing the largest response, despite significant differences in the ridge width/gap utilised. Nevertheless, our model predicts that the strongest response occurs when gap and width are equal, at $(W, G) \approx (0.67, 0.67)$, i.e. for relatively wide ridges. 

For constant $G$ or $W$ equal to 0.67, significant amplification is observed when varying $W$ or $G$, respectively, along the two orthogonal red dashed lines in figure \ref{map_re}. Along these directions, one additional local peak is clearly visible at spacing $S\approx2.8$, but several other (weaker) peaks occur at higher gaps or widths, at integer multiples of the optimal width $W \approx 0.67$. It is anticipated that these further peaks correspond to configurations with strong tertiary, quaternary of high-order structures \citep{Hwanglee2018} above/within the ridge/trough, confirming the conceptual model of \citep{medjnoun_vanderwel_ganapathisubramani_2020}. Nonetheless, further increasing the width (respectively, the gap) at constant gap (respectively, $W$) does not produce major changes in the strength of the response. These configurations tend asymptotically to the isolated ridge (respectively gap) state, where the interaction between flow structures generated by adjacent ridges (respectively, gaps) can be neglected and the response is constant, regardless of the measure utilised. 

A further important observation is that the response shows a symmetry with respect to the line $G = W$. The symmetry arises from the linear nature of the present analysis. For any surface configuration $(W, G)$, the flow topology in the trough is identical but with opposite flow direction to that on the ridge when $G$ and $W$ are swapped. The symmetry of the problem implies that the conceptual model developed by \citet{medjnoun_vanderwel_ganapathisubramani_2020} speculating on the formation of tertiary structures over wide ridges can also be employed to describe the formation of tertiary structures in wide troughs induced by ``virtual roughness element'' as proposed by \citet{vanderwel2015}.

Finally, the implication of the response maps of figure \ref{map_re} is that, despite the spacing $S$ is a relevant length scale to characterize secondary flows, two surface parameters are required to characterize in a complete manner the strength of secondary currents. While many choices are possible, e.g. $S$ and $W$ as in \citet{Hwanglee2018,vanderwel2015,medjnoun_vanderwel_ganapathisubramani_2020} or $S/W$ and $S$ as in \citet{castro2020}, using $G$ and $W$ is particularly convenient as i) the response has a symmetry with respect to the line $G=W$ and ii) these two parameters have similar roles when the flow organisation is considered, as we discuss in the next section.

 {A comparison with harmonic wall modulations is now performed to highlight similarities and differences between the two types of heterogeneity. To enable a direct comparison, we use the period $\lambda_3$ for harmonic walls and the spanwise spacing $S$ for rectangular ridges. Hence, for the latter case, where two geometrical parameters are strictly necessary, we restrict the analysis to specific configurations at duty cycle $DC=0.25,0.5$ and for a fixed width $W=0.67$, corresponding to the special identified by the additional lines in figure \ref{map_re}. Due to the symmetry of response discussed above, cases at duty cycles $DC^\prime$ greater than 0.5 have the same kinetic energy density and streamfunction peak of a cases at $DC=1 - DC^\prime$. For both types of heterogeneity, the same infinitesimal peak-to-peak modulation amplitude is used. 
The major similarity between the two types of heterogeneity is that the secondary structures have maximum intensity at a similar spanwise period, regardless of the quantity considered. For sinusoidal walls, the peak occurs at  $\lambda_3 \approx 1.54$ while for rectangular ridges the peak is observed at $S \approx 1.34$. The effect of the duty cycle is only moderate, due to the orientation of the contours of the kinetic energy density in figure \ref{map_re}. The small difference between the two types of heterogeneity can be attributed to the fact that the first harmonic mode of the expansion (2.4) for the rectangular ridge geometry given by equation (4.1) is the largest and hence provides the dominant contribution to the overall response.
However, the peak strength depends on the duty cycle. For instance, at $DC=0.25$, the strength of secondary flows is 75\% less intense than the vortices generated for the same spacing at $DC=0.5$. As discussed later on in section \ref{sec:topology_secondary_flows}, secondary vortices do not have enough lateral space to develop at a sufficient strength when the ridges are too narrow, i.e. for small duty cycles. Interestingly, the peak value observed in figure \ref{fig:review} for rectangular ridges can be larger than what is observed for sinusoidal walls, provided the duty cycle is selected appropriately to intersect the high amplification region in the map of figure \ref{map_re}. Within the present linear framework, this effect can be explained to be the result of the constructive interference of the individual flow responses at all wavenumber modes defining the rectangular ridge geometry.
A second major difference is that the response for rectangular ridges can exhibit a second peak at a larger spanwise period depending on the duty cycle. In particular, a peak at $S\approx2.5$ can be observed at $DC=0.25$ and for $W=0.67$, while for $DC=0.5$, the second peak is shifted to a higher spacing ($S\approx4$), associated to configurations at higher $W$ and $G$. It is anticipated that these secondary peaks correspond to ridge configurations where tertiary flows, developing at the center of the ridge (or troughs), have the maximum strength.

\begin{figure}
    \centering
    \includegraphics[width=\textwidth]{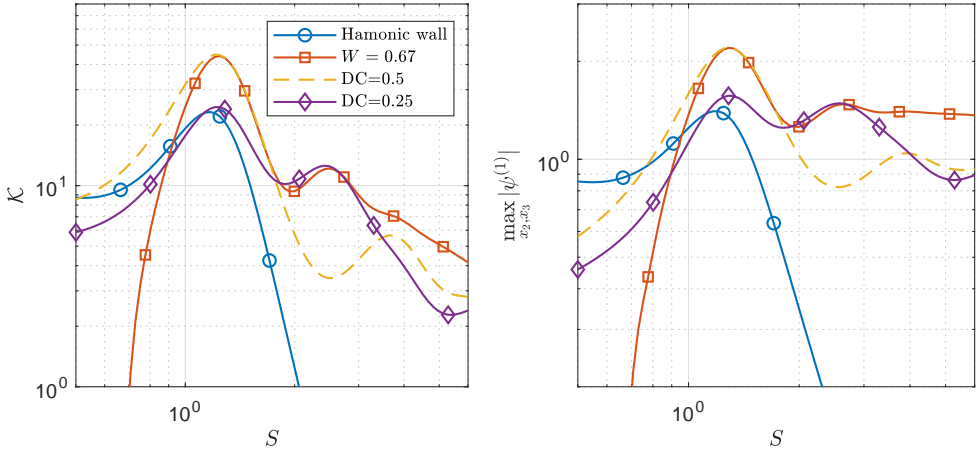}
     \caption{Comparison of the kinetic energy density $\mathcal{K}$ (left panel) and the maximum of the streamfunction $\max_{x_2,x_3} |\psi^{(1)}|$ (right panel) at $Re_\tau=5200$ as a function of the periodicity $S$.
    For the rectangular ridges, the quantities are obtained for constant $W=0.67$ and for $DC=0.25$ and $0.5$. 
    }
    \label{fig:review}
\end{figure}
}

\subsection{Topology of secondary flows}
\label{sec:topology_secondary_flows}
\begin{figure}
    \centerline{
    \includegraphics[width=\textwidth]{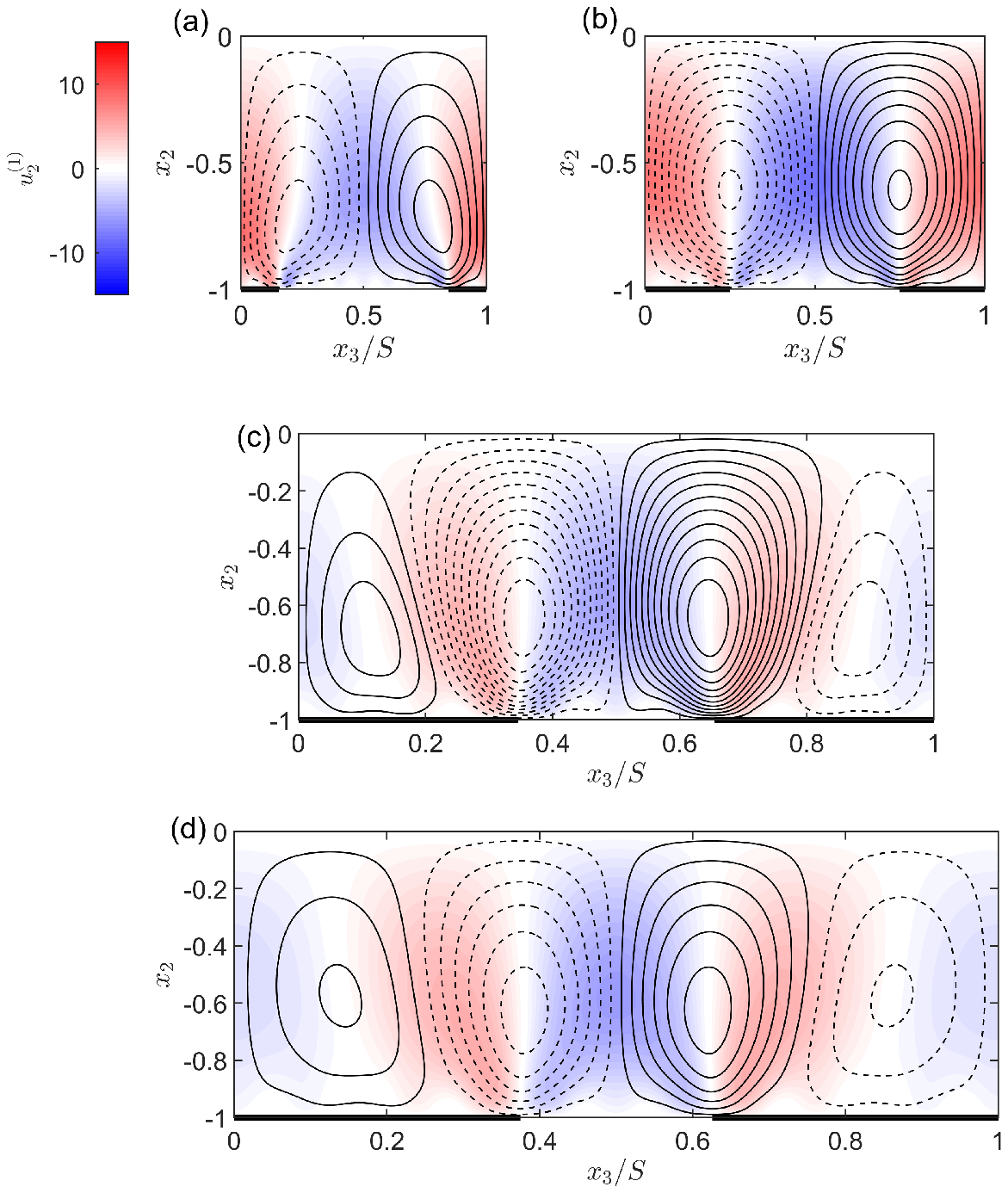}
    }
    \caption{Flow organization for $G=0.67$ and width $W=0.3$ (a), 0.67 (b), 1.5 (c) and 2 (d). Results for $\Rey_{\tau}=5200$ are shown. Contours of the perturbation streamfunction $\psi^{(1)}$ between -2 and 2 are shown. The dashed lines indicate negative streamfunction values. The colour map of the perturbation wall-normal velocity component $u_2^{(1)}$ is also reported in the lower half of the channel. The ridges are sketched at the bottom line using bold lines. Note that the ridges are centered at $x_3=0$ and 1 and the fields are spanwise periodic.}
    \label{isolineconstantsw}
\end{figure}

Based on the symmetry highlighted from the response maps, we now show how the parameters $W$ and $G$ affect the organization of secondary flows. We consider flows at $\Rey_\tau=5200$, at which the response has saturated to its high-Re asymptotic state. In figure \ref{isolineconstantsw}, contours of the perturbation streamfunction are reported together with colour maps of the wall-normal velocity perturbation for configurations at constant gap $G=0.67$ and at varying $W = 0.3, 0.67, 1.5$ and $2$ (see triangles in figure \ref{map_re}). The black lines at $x_2 = -1$ define the locations of the ridges. Note that the fields are spanwise periodic and only half of the ridge is shown, as the ridges are centred at $x_3 = 0$. 
 
Starting from $W=0.3$, the linear model predicts counter-rotating vortical structures elongated in the wall-normal direction and occupying the entire half-width of the channel. These structures are locked in proximity of the ridge edges where the surface discontinuity acts as a strong source term. A downwash inside the troughs and an upwash above the ridges in proximity of the edge is observed. The  maximum intensity of these vertical motions at $W=0.67$ is about 15$u_\tau$, per unit of ridge height. This means that for a peak-to-peak ridge height of $0.09$ (in units of the boundary layer thickness) as in case HS6 \citep{medjnoun_vanderwel_ganapathisubramani_2020} for $\Rey_{\tau}=3239$, the predicted peak vertical velocity is $3\%$ of the bulk velocity, which agrees with the experimental data ($2\%$).
 {However, the comparison between the case HS6 of \citet{medjnoun_vanderwel_ganapathisubramani_2020} and the present model can only be qualitative because of structural differences between channel flows and boundary layers. In particular, wall-normal secondary motions in a boundary layer are not confined or blocked by the upper wall (or symmetry plane), but can develop freely. This effect is likely more important when the secondary structures have size comparable to the shear layer thickness, and secondary motions can reach the outer edge of the turbulent shear flow.}
For a short $W=0.3$, panel (a), the vortical structures compete for the available space over the ridge, push each other towards the gap centre and are highly elongated in the vertical direction. For $W = 0.67$, panel (b), the vortices can now fully extend towards the ridge centre. For $W = 1.5$, panel (c), there is sufficient space over the ridge for tertiary flows to emerge in the region immediately above the ridge. In this condition, downwash is observed over the ridge centre {.} This, however, is associated to a reduction in the strength of the upwash in the vicinity of the edges, strongest at $W=0.67$. Tertiary vortical structures are initially weak but gain strength at $W\approx2.1$, where they can fully extend to the channel mid-plane. The strength of the downwash velocity at the ridge centre for $W=2.1$ relative to the downwash velocity over the gap is significant. This is likely exacerbated by confinement effects in the channel, in which the spanwise-averaged vertical mass transport operated by secondary currents is necessarily zero. In boundary layers, no such constraint would exist. Although not shown here, for $W > 3.5$ a further reorganization is observed, where weak quaternary vortical structures emerge near the ridge centre ($x_3=0$), producing a weak upwash motion. 

One important remark is that the present linearised model does not capture correctly flow features observed in the immediate vicinity of the ridge such as, for instance, recirculation regions induced by strong spanwise motions over the ridge top, frequently observed in direct numerical simulations \citep{Hwanglee2018, castro2020}. The wall-normal extent of these regions is a) strongly influenced by the ridge geometry (rectangular, circular, etc) and b) likely scaling with the ridge height, which is always finite in experiments and simulations. In the present linear model, the ridge height is infinitesimal and only large-scale flow features developing far away from the surface are likely to be captured correctly. Localised near-wall effects produced by a finite ridge height and contributing less prominently to the alteration of vertical transport phenomena are unlikely to be accounted for.  {Nevertheless, the model predicts structures with similar characteristics to those observed in many other other studies, where the ridges protrude into the log region. It could be argued that, for a shallow surface modulation, all the mean flow quantities (e.g. the Reynolds stresses) develop in the wall-normal direction according to the same law of the wall, as if the wall was flat. The lateral variation of the origin of these profiles produced by the modulation then produces at any distance from the wall spanwise gradients of the Reynolds stresses, i.e. the required source terms in the streamwise vorticity equation \eqref{eq:linear_equations}. This mechanism might be at play regardless of the height of the modulation, although for large protrusions other mechanism became relevant, e.g. the wall-normal deflection of spanwise velocity fluctuations (see e.g. Hwang \& Lee 2018). }

\begin{figure}
   \centerline{
    \includegraphics[width=0.96\textwidth]{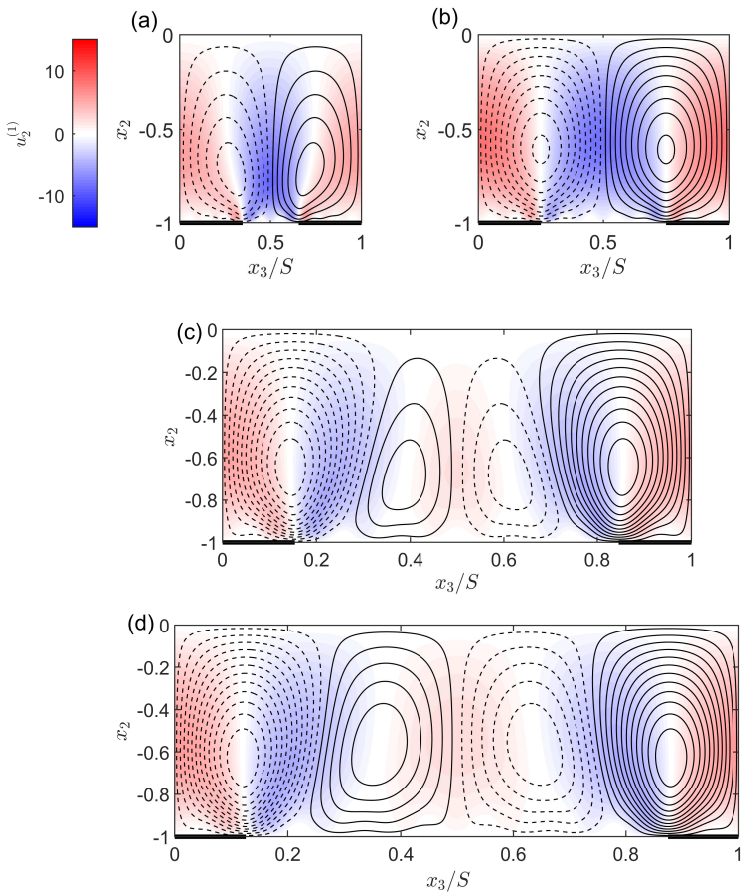}}
    \caption{Flow organization for $W=0.67$ and gap $G=0.3$ (a), 0.67 (b), 1.5 (c) and 2 (d). Results for $\Rey_{\tau}=5200$ are shown. Contours of the perturbation streamfunction $\psi^{(1)}$ between -2 and 2 are shown. The dashed lines indicate negative streamfunction values. The colour map of the perturbation wall-normal velocity component $u_2^{(1)}$ is also reported is also reported in the lower half of the channel. The ridges are sketched at the bottom line using bold lines.  Note that the ridges are centered at $x_3=0$ and 1 and the fields are spanwise periodic.}
    \label{isolineconstantw}
\end{figure}

For completeness, the evolution of the flow organisation for a constant $W=0.67$ as the gap $G$ increases is shown in figure \ref{isolineconstantw}. These configurations correspond to the squares in figure \ref{map_re}, and parallel the configurations shown in figure \ref{isolineconstantsw}. For $G=0.3$, panel (a), the vortical structures compete for the available space over the gap and push each other away towards the ridge. As the gap is further increased to $G=1.5$ and then 2, tertiary structures form in the centre of the trough producing vertical velocities weaker than the velocity induced by the secondary structures over the ridge. As anticipated, this behaviour was described by \citet{vanderwel2015}, who observed that, when the spacing is large enough, an additional upwelling motion is generated at the centre of the trough as if a ``virtual'' ridge element was placed between physical ridges.

 In conclusion, the secondary structures shown in \ref{isolineconstantsw} and \ref{isolineconstantw}  and predicted by the present model are similar in size and organisation, especially in the region closest to the wall, to the secondary currents observed over strip-type roughness both numerically \citep{anderson2015,chung2018} and experimentally \citep{hinze1967,hinze1973,mejia2010,mejia2013,barros2014,anderson2015,bai2018,forooghi,wangsawijaya2020}. The similarity may be due to the fact that for strip-type heterogeneity, the wall-normal extent of the surface perturbation is localised, loosely speaking, within the roughness height. When scaled in outer units, this height is often much smaller than the typical (and finite) ridge height used in previous work. However, the key difference is that the direction of rotation of secondary structures over strip-type roughness is opposite to what is observed for ridge-type heterogeneity and a downwash motion is predicted over the high roughness region \citep{stroh2019,schafer2022}.
The strong similarity suggests that the present linearised framework may also be adapted to study secondary currents produced by other types of complex surfaces, where the lateral variation of the surface attributes may be modelled accurately by suitable boundary conditions that capture the physics of the flow-surface interaction. In addition to the strip-type roughness case just discussed, examples of such surfaces include superhydrophobic surfaces \citep{busse2012,turk2014,stroh2016}, porous surfaces \citep{Bottaro:2019jv, Rosti:2018cq, AbderrahamanElena:2017jf, Efstathiou:2018fh} or surface lateral variations of the heat flux \citep{salesky_calaf_anderson_2022}.

 {
\subsection{Velocity profiles over rectangular ridges}
For a better characterization of the flow structures developing over the ridges, the wall-normal velocity profile at five different locations between the left edge of the ridge and its centre are reported in figure \ref{fig:tanwave_ver5}. The sketch at the bottom of the figure illustrates the location where profiles are extracted. The velocity profiles are obtained at $\Rey_\tau=5200$ and for $DC=0.5$. The ridge width $W$ varies from $0.3$ in panel (a) to $2$ in panel (d). For the two smaller widths,  the velocity is always positive corresponding to the upwash region of figure \ref{isolineconstantsw}(a, b). For the optimal configuration $W=0.67$, where the vortical structures fit the available space without significant lateral distortion, the velocity above the ridge edge is small. This contrasts with experimental/numerical observations (e.g. \cite{medjnoun_vanderwel_ganapathisubramani_2020}, where intense upwards motions are often observed at the ridge edge. This might be the result of the finite ridge height in these cases, which produces a wall-normal deflection of the  spanwise velocity fluctuations. The present linear model, with infinitesimal ridges, does not capture this deflection although it does predict an intense spanwise motion at the ridge edge.
Tertiary flows occur for $W = 1.5$ and $2$, where the $u_2^{(1)}$ profile at the centre of the ridges (light red) displays a negative  value.  
However, the negative peak is about 50\% less intense than the positive one, although the size of the region interested by the wall normal motion is similar (panels c and d). The intensity of the tertiary flows slightly increases with $W$ while the secondary flows appear to be unaffected. This suggests a saturation of the secondary flows, while higher order structures occur at the ridge centre. 
\begin{figure}
    \centering
    \includegraphics[width=\textwidth]{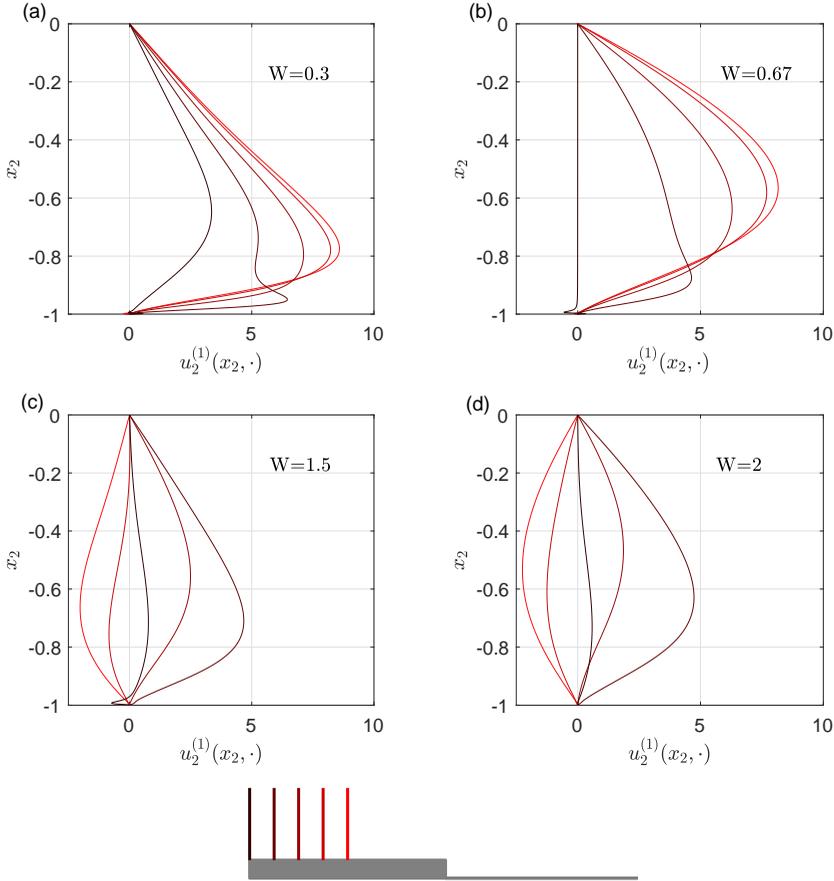}
    \caption{Wall-normal velocity profiles at different locations along the ridge. The Reynolds number is $\Rey_\tau=5200$ and duty cycle is $DC=0.5$. The width W varies from $0.3$ (panel a) to 2 (panel d). The profile locations are also reported in the sketch at the bottom of the figure using a different colour gradation.}
    \label{fig:tanwave_ver5}
\end{figure}}

\begin{figure}
    \center
    \includegraphics[width=0.74\textwidth]{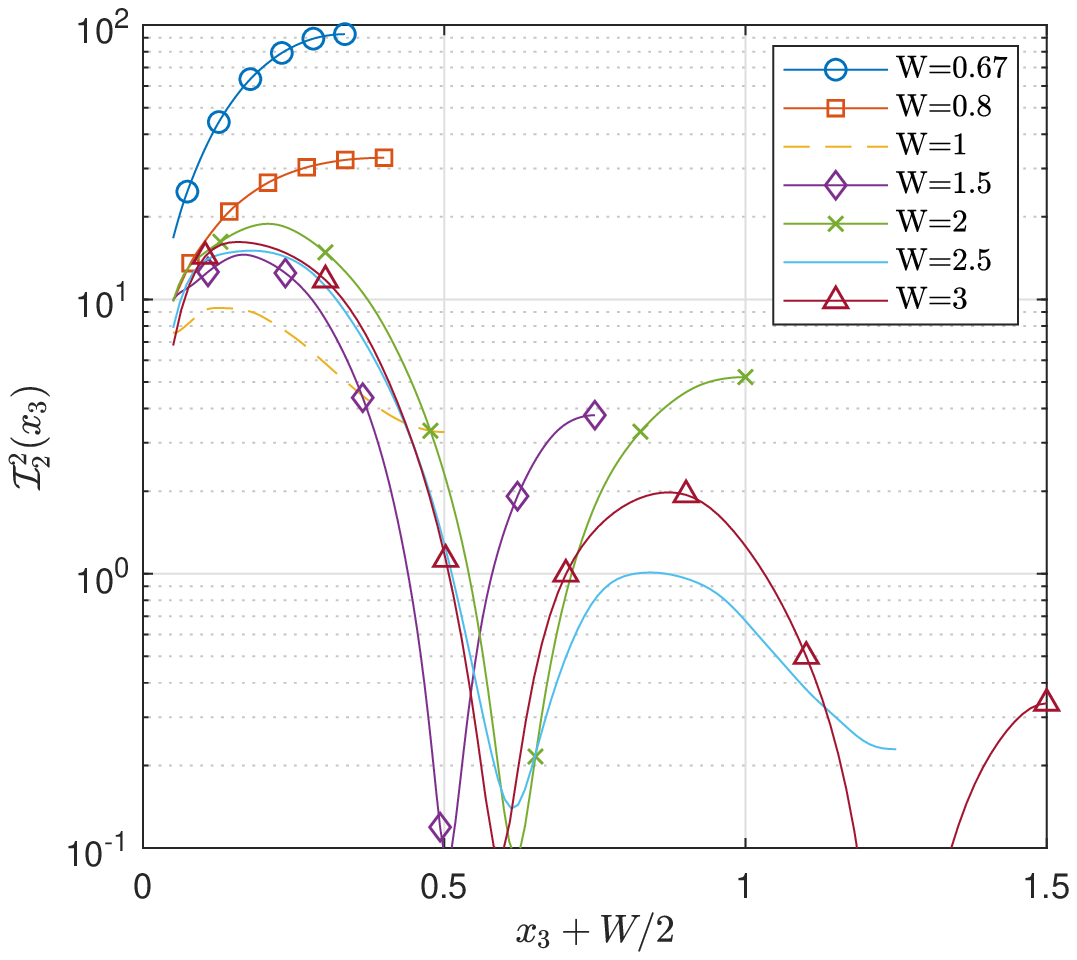}
    \caption{ {The quantity $\mathcal{I}_2^2(x_3)$ for $\Rey_\tau=5200$ and duty cycle $DC=0.5$. The ridge width $W$ varies from 0.67 to 3. In particular, ---$\!\!\circ\!\!\!$--- $W=0.67$, ---$\!\!\square\!\!\!$--- $W=0.8$,  $---$ $W=1$, ---$\!\!\Diamond\!\!\!$--- $W=1.5$, ---$\!\!X \!\!\!$--- $W=2$, ------ $W=2.5$ and  }}
    \label{fig:profile_integral}
\end{figure}
 {
To better characterize the intensity and direction of the secondary flows as a function of the spanwise location, the quantity  
\begin{equation}
    \mathcal{I}_2^p(x_3)=\int_{-1}^{0} u_2^{(1)^p}(x_2,x_3)\; \mathrm{d}x_2
\end{equation} is now introduced. We first discuss the case for $p=2$. Results are reported in figure \ref{fig:profile_integral} for $DC=0.5$ and $W$ varing from 0.67 to 3. To align all profiles, we use the spanwise coordinate $x_3+W/2$, so that the ridge edge is always located at $0$ while the ridge centre corresponds to $W/2$. 
For the smaller widths, the quantity $\mathcal{I}_2^2(x_3)$ is quite intense and a single peak produced by the secondary structures locked on the ridge edges is observed,  approximately at the ridge centre.
Increasing the ridge width to $W=1.5$, the quantity $\mathcal{I}_2^2(x_3)$ shows two peaks: the first in proximity to the ridge edge (with smaller magnitude than at the optimal width $W=0.67$) and the second at the ridge centre, characterising the strength of tertiary flows.
When the width is further increased, secondary flows develop fully and only moderate effects on their strength near the ridge edge is observed. Major differences are still observed for tertiary flows at the ridge centre, although the expectation is that such differences would eventually vanish as the ridge width is increased further.}

\subsection{Analysis of  {the wall-normal velocity direction over the ridge centre}}
\begin{figure}
    \center
    \includegraphics[width=0.9\textwidth]{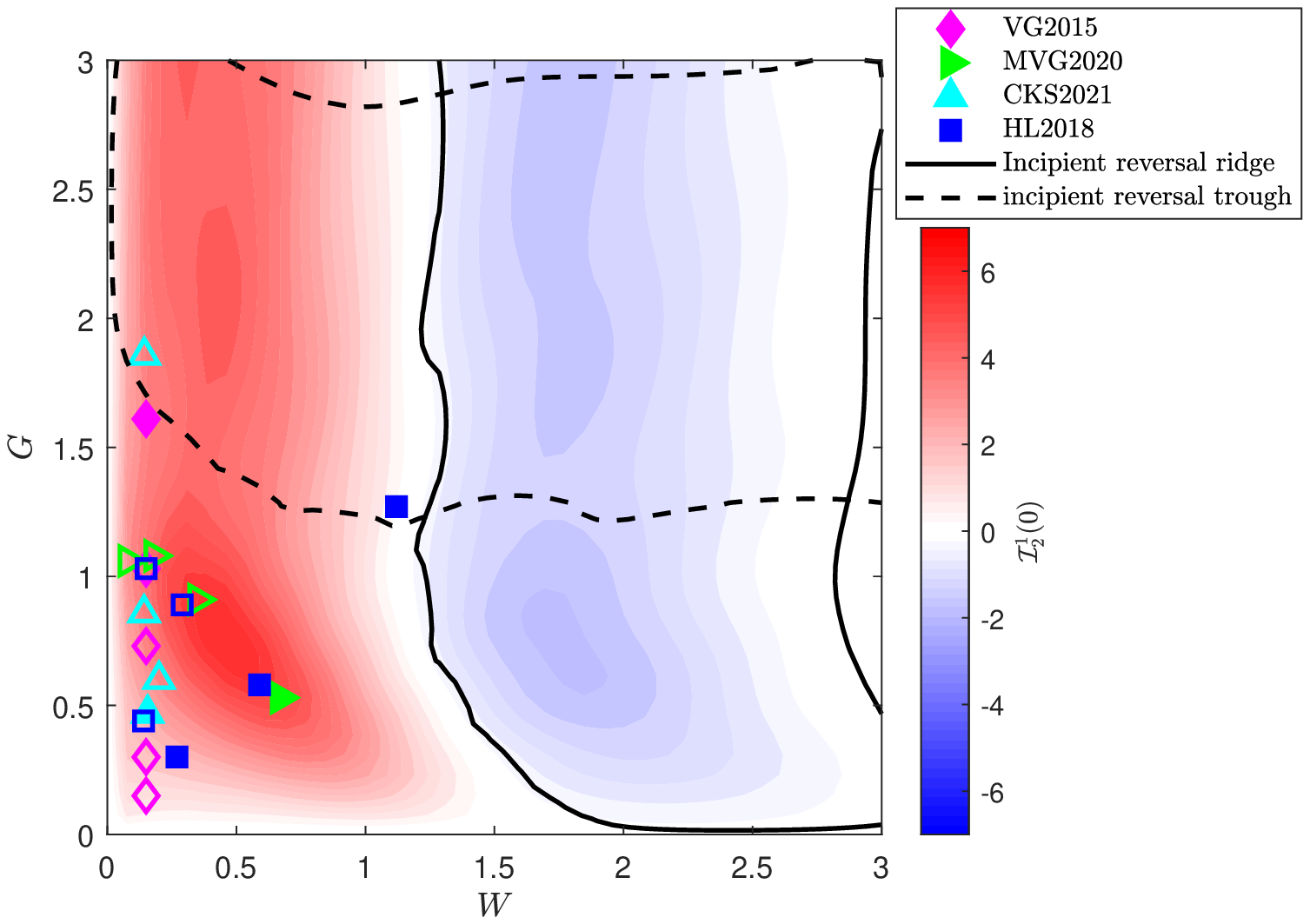}
    \caption{Colour map of the quantity  {$\mathcal{I}_2^{1}(0)$} as a function of the gap $G$ and width $W$, for $Re_\tau = 5200$. Configurations studied in the recent literature are denoted by symbols (VG2015 for \citet{vanderwel2015}, MVG2020 for \citet{medjnoun_vanderwel_ganapathisubramani_2020}, CKS2021 for \citet{castro2020} and HL2018 for \citet{Hwanglee2018}). Closed symbols denote configurations where  {downwash over the ridge} has been observed. The black lines delimit the regions where the linear model predicts incipient  {change of flow direction} at the midpoint over the ridge (solid line) and at the centre of the trough (dashed line).}
    \label{fig:squarewave_vel}
\end{figure}
To better visualize the region of parameter space where the present linearised model predicts a large-scale  {change in the flow direction above} the centre of the ridges, we  {use the quantity $\mathcal{I}_2^1(0)$} to quantify the average, or ``bulk'', wall-normal flow direction at $x_3 = 0$, as a function of the gap $G$ and the width $W$. Results are reported in figure \ref{fig:squarewave_vel}. The linearised model indicates that the bulk wall-normal velocity becomes negative for $W \gtrsim 1.2$, with a moderate effect of the gap. The maximum average velocity occurs for $W\approx0.5, G\approx0.75$, indicating that optimising the intensity of secondary currents using the strength of the average wall-normal velocity yields narrower ridges than what suggested by using the integral perturbation energy or the streamfunction peak. The bulk velocity turns positive again for $W \gtrsim 2.8$ when the ridge is wide enough to support the formation of quaternary structures. The quantity  {$\mathcal{I}_2^{1}(0)$} alone, however, might not be sufficient to capture  {the change in flow direction} that is often observed in the proximity of the obstacle \citep{castro2020}. The onset of  {this change} is thus also indicated in the figure, by tracing the set of points (solid black line) in parameter space where the wall-normal velocity at the centre of the ridge first changes sign. Due to the aforementioned symmetries, large-scale or incipient  {change in flow direction} in the troughs, observed e.g. by \citet{vanderwel2015} can be characterised by swapping the role of $G$ and $W$ and inverting the sign of  {$\mathcal{I}_2^{1}(0)$} (computed at $x_3 = S/2$, in the centre of the trough). The region where  {a change in flow direction} is predicted in the troughs by the present model is shown as a dashed black line. The model predicts that the difference between the average and incipient  {change of flow direction} is minimal. However, this difference might be more pronounced for finite height ridges, where the flow topology near the ridge is more complicated than what can be captured by the present linear model.

Data from recent numerical and experimental investigations that have considered streamwise rectangular ridges are also reported in figure \ref{fig:squarewave_vel}.  {As a note of caution, most of these cases (with the exception of \citet{castro2020} who considered channel flows) are extracted from studies of secondary flows in boundary layers. As discussed previously, secondary flow structures originating in different flow types might display significant topological differences and the following analysis should be regarded as qualitative.} Interestingly, a large fraction of experiments and numerical simulations available in literature is focused on the region of narrow width, relatively far from the optimal configuration predicted by the present model. In the figure, closed symbols denote configurations where  {a large scale change in the flow direction} (and not simply in the neighbourhood of the ridge) was observed above the ridge or in the trough.
These are the case HS6 from \citet{medjnoun_vanderwel_ganapathisubramani_2018} and P24S12 from \citet{Hwanglee2018}, where a downwelling motion is observed above the ridge at large distance from the wall, and $S/\delta=1.76$ from \citet{vanderwel2015}, where upwelling is measured in the trough. For case HS6, the present model predicts a positive net wall-normal velocity, in contrast to experimental evidence. Inspection of the velocity field for this case in \citet{medjnoun_vanderwel_ganapathisubramani_2018} shows that the Reynolds-averaged vortical structures are smaller in size (in both directions) and less coherent than what predicted by the present model. In turn, this would increase the space available for fluid to reverse its direction. 
 {This difference might be due to the different flow type (boundary layer in \citet{medjnoun_vanderwel_ganapathisubramani_2018} and channel flow in the present work) or to the finite ridge height in experiments.}

\section{Conclusions} \label{conclusions}
A rapid tool for the prediction of secondary currents developing in turbulent channels with streamwise-independent surface modulations has been presented. The approach is based on the linearisation of the steady Reynolds-averaged Navier-Stokes equations, coupled to the Spalart-Allmaras equation for the transport of the turbulent eddy viscosity.  The linearisation of these equations is based on the assumption that the surface modulation is small when compared to any relevant geometric or physical length scale. The influence of the surface modulation is then modelled using inhomogeneous boundary conditions for the streamwise velocity component and the turbulent eddy viscosity. Because of the linearity, the superposition principle applies and the flow response induced by an arbitrary surface with spectrally-complex topography can be obtained by appropriately combining the elementary responses to harmonic modulations at each spanwise wavelength.

The computational efficiency of the tool allows large parameter spaces characterising complex surfaces to be explored at little cost. In this paper, two canonical surface configurations are studied, namely, harmonic modulations and rectangular ridges. For harmonic modulations, characterised by a single spanwise length scale, the wavelength $\lambda_3$, the turbulent shear flow is found to have the largest response at two spanwise wavelengths, scaling in inner and outer units, respectively. The outer peak is found at $\lambda_3 \approx 1.54$, in units of the half-channel width, and corresponds to large-scale secondary vortices that occupy the entire half-channel width. These produce an upwelling motion over the crests and a downwelling motion over the troughs, with no tertiary vorticity observed. The inner peak, of much lower intensity, is found at $\lambda_3^+ \approx 45$ and corresponds to small scale vortices extending by about 30 viscous units in the wall normal direction. The presence of two peaks mirrors the results of \emph{transient} growth analysis in turbulent channels by \citet{delalamo2006} and \citet{pujals2009} and suggests that surface topography modulation of the right spanwise length scale can excite a strong, steady response by leveraging amplification mechanisms intrinsic to the turbulent shear flow. However, a major difference with the optimal structures found by these works is that the strength of the \emph{steady} response to surface modulations predicted by the present tool becomes asymptotically Reynolds-number-independent when the cross-plane velocities are scaled with the friction velocity. Fundamentally, this is due to the Spalart-Allmaras transport model utilised in this work, designed to produce the law of the wall and in which the turbulent eddy viscosity (and the Reynolds stresses driving secondary currents) become, asymptotically, Reynolds-number-independent.

For rectangular ridges, the present model suggests that a) both the ridge width $W$ and the gap between ridges $G$ are key parameters to quantify the response and that b) the analysis is more revealing when these two parameters are used and not other combinations previously used in the literature. More importantly, the largest response is found at a symmetric configuration where $W=G=0.67$, i.e. a rather large ridge width for a spanwise spacing of $S=G+W \approx 1.34$. For other ridge configurations, the secondary vortices compete for the available space with structures developing on the same ridge or over neighbouring ridges or are weakened by tertiary structures appear at large gaps or widths.

 {It is important to mention that the proposed approach has its limitations and it should not be seen as a replacement for DNS or experiments to obtain precise quantitative predictions. Secondary currents have been shown to display highly unsteady behaviour \citep{vanderwel2019} and meander in the 
spanwise direction \citep{hutchins2007,kevin2019,zampiron2020}. The present model assumes steady, streamwise-independent perturbations, and cannot fully capture any of these phenomena. Secondly, it is likely that surfaces characterised by prominent ridges, or surfaces with rapid lateral variations of the geometry (i.e. surfaces with sharp corners or with large lateral slope) cannot be satisfactorily modelled using linearised boundary conditions. For example, several authors (e.g. \citet{Hwanglee2018}) have observed that the wall-normal deflection of spanwise velocity fluctuations operated by the vertical sides of rectangular ridges is an important mechanism for the generation of secondary currents. This effect may be important, especially when the ridges protrude significantly in the log-layer and convective effects result in complex flow topologies, as in \citet{castro2020}. This mechanism is clearly not accounted for in the present model, where the ridge height is infinitesimal. Nevertheless, the model does produce secondary structures that resemble those observed in DNS or experiments. Most importantly, it correctly predicts the spanwise ridge spacing at which peak strength is achieved, in agreement with previous observations. This suggest that while different generation mechanisms might be at play, the linearised Navier-Stokes operator still provides an adequate description of the response to an external forcing.}

With appropriate modelling assumptions, the present approach would also enable a rapid exploration of the vast parameter space characterizing other surface heterogeneities that have been recently considered in the literature, e.g. strip-type roughness \citep{willigham2014, anderson2015, chung2018}, super-hydrophobic surfaces (e.g. \citet{turk2014,stroh2016}) or combinations of topography and roughness, as in e.g.  {\citet{stroh2019,schafer2022}. However, modelling the physics of the flow-surface interaction within a linear framework is less straightforward than modelling a modulation of the surface height, as in the present case. These configurations are currently being considered and results will be reported in future work}.\\

\textbf{Data access statement.} All data supporting this study are openly available from the University of Southampton repository at https://doi.org/10.5258/SOTON/D2218.\\

\textbf{Declaration of Interests}. The authors report no conflict of interest.\\

\appendix
\section{Linearization of the normalised rotation tensor}\label{appendix:qcr}
Expression of the normalised  rotation tensor at order zero and order one are reported
 \begin{align}
    O^{(0)}&=\left[\begin{matrix} 0 & sign(\Gamma) & 0 \\
    -sign(\Gamma) & 0& 0\\
    0 &0 &0	
    \end{matrix}\right], \\
   O^{(1)}&=\left[\begin{matrix} 0 & 0 & \frac{sign(\Gamma)}{\Gamma} \frac{\partial u_{1}^{(1)}}{\partial x_{3}} \\
    0 & 0& \frac{sign(\Gamma)}{\Gamma} \left(\frac{\partial u_{2}^{(1)}}{\partial x_{3}}-\frac{\partial u_{3}^{(1)}}{\partial x_{2}} \right)\\
    -\frac{sign(\Gamma)}{\Gamma} \frac{\partial u_{1}^{(1)}}{\partial x_{3}}&-\frac{sign(\Gamma)}{\Gamma} \left(\frac{\partial u_{2}^{(1)}}{\partial x_{3}}-\frac{\partial u_{3}^{(1)}}{\partial x_{2}} \right) &0	
    \end{matrix}\right],
   \end{align}
   where $\Gamma$ is the zero-order streamwise velocity wall-normal gradient and $sign$ is the sign function.

\section{Terms of the linearized SA model} \label{app:termini}
In this section, additional terms appearing in the linearised Spalart-Allmaras transport equation (\ref{SAfirstorder1}) are reported. Firstly, terms in (\ref{eq:nut1}) are \begin{equation}
\tilde{f}_{v1}= 3 \Rey_\tau^3 c_{v1}^3 \displaystyle \frac{\tilde{\nu}^{(0)^2}}{\left(\Rey_\tau^3 \tilde{\nu}^{(0)^3}+c_{v1}^3\right)^2\tilde{\nu}^{(1)}}.
\end{equation}
Similarly, the source term $\tilde{\mathcal{S}}$ can be written as the sum of a zero order and first order contributions, too. Thus, 
\begin{equation}
    \tilde{\mathcal{S}}=\tilde{\mathcal{S}}^{(0)}+\epsilon \tilde{\mathcal{S}}^{(1)},
    \label{eqn:S_definition}
\end{equation}
where the zero order function $\tilde{\mathcal{S}}^{(0)}$ is readily obtained from its nonlinear definition.
Furthermore, the first order $\tilde{\mathcal{S}}^{(1)}$ is here decomposed into $\tilde{\mathcal{S}}^{(1)}=\tilde{\mathcal{S}}_1 \tilde{\nu}^{(1)}+ \tilde{\mathcal{S}}_2 \frac{\partial u_1^{(1)}}{\partial x_2}+\tilde{\mathcal{S}}_3 d^{(1)}$ where
\begin{subeqnarray}
\tilde{\mathcal{S}}_1 &=& \displaystyle \frac{f_{v2}^{(0)}}{k^2 d^{(0)^2}}+ \displaystyle \frac{\tilde{\nu}^{(0)}}{k^2 d^{(0)^2} f_{v2}^{(1)}},\\
\tilde{\mathcal{S}}_2 &=& sign{(\Gamma)},\\
\tilde{\mathcal{S}}_3 &=& -2 \frac{\tilde{\nu}_t f_{v2}^{(0)}}{k^2 d^{(0)^3}}.
\end{subeqnarray}
Similarly, the function expanded in $f_{v2}=f_{v2}^{(0)}+ \epsilon \tilde{f}_{v2}^{(1)}$ where
\begin{equation}
\tilde{f}_{v2} = - \Rey_\tau \displaystyle \frac{c_{v1}^6 \tilde{\nu}^{(0)^6}+ \Rey_\tau^3 c_{v1}^3 \tilde{\nu}^{(0)^3} (2-3 \Rey_\tau \tilde{\nu}^{(0)}) }{\left[c_{v1}^3+\Rey_\tau^3 \tilde{\nu}^{(0)^3}(1+\Rey_\tau \tilde{\nu}^{(0)}) \right]^2} \tilde{\nu}^{(1)}.
\end{equation}
Finally, the remaining terms of the Spalart-Allmaras model can be written as 
\begin{subeqnarray}
    r&=&r^{(0)}+\epsilon\left(r_1 \tilde{\nu}^{(1)}+r_2 \frac{\partial u_1^{(1)}}{\partial x_2}+r_3 d^{(1)}\right),\\
    g&=&g^{(0)}+\epsilon\left(g_1 \tilde{\nu}^{(1)}+g_2 \frac{\partial u_1^{(1)}}{\partial x_2}+g_3 d^{(1)}\right),\\
    f_w&=&f_{w}^{(0)}+\epsilon\left(f_{w_1} \tilde{\nu}^{(1)}+f_{w_2} \frac{\partial u_1^{(1)}}{\partial x_2}+f_{w_3} d^{(1)}\right).
\end{subeqnarray}
where
\begin{subeqnarray}
r_1 &=& \frac{\tilde{\mathcal{S}}^{(0)} d^{(0)}-\tilde{\nu}^{(0)} \tilde{\mathcal{S}}_2 d^{(0)}}{\tilde{\mathcal{S}}^{(0)^2}k^2 d^{(0)^3}},\\
r_2 &=&\frac{-\tilde{\nu}^{(0)} \tilde{\mathcal{S}}_1 d^{(0)}}{\tilde{\mathcal{S}}^{(0)^2}k^2 d^{(0)^3}},\\
r_3 &=& \frac{- \tilde{\nu}^{(0)} \tilde{\mathcal{S}}_3 d^{(0)}- 2 \tilde{\nu}^{(0)} \tilde{\mathcal{S}}^{(0)}}{\tilde{\mathcal{S}}^{(0)^2}k^2 d^{(0)^3}}.
\end{subeqnarray}
Similarly, for $g$ we have 
\begin{subeqnarray}
g_1&=& \left(\left(6 r^{(0)^5}-1\right) c_{w2}+1\right) r_1,\\
g_2&=& \left(\left(6 r^{(0)^5}-1\right) c_{w2}+1\right) r_2,\\
g_3&=& \left(\left(6 r^{(0)^5}-1\right) c_{w2}+1\right) r_3,\\
\end{subeqnarray}
while for $f_{w}$ we have 
\begin{subeqnarray}
f_{w_1}=\displaystyle \frac{c_{w3}^6}{c_{w3}^6+1} \displaystyle \left( \frac{c_{w3}^6+1}{g^{(0)^6}+c_{w3}^6} \right)^{\frac{7}{6}} g_1,\\
f_{w_2}=\displaystyle \frac{c_{w3}^6}{c_{w3}^6+1} \displaystyle \left( \frac{c_{w3}^6+1}{g^{(0)^6}+c_{w3}^6} \right)^{\frac{7}{6}} g_2,\\
f_{w_3}=\displaystyle \frac{c_{w3}^6}{c_{w3}^6+1} \displaystyle \left( \frac{c_{w3}^6+1}{g^{(0)^6}+c_{w3}^6} \right)^{\frac{7}{6}} g_3.
\end{subeqnarray}

\bibliographystyle{jfm}
\bibliography{jfm-instructions}

\end{document}